\documentclass[sigconf, nonacm, pdfa]{acmart}

\usepackage{hyperref}
\usepackage{cleveref}
\usepackage{tikz}
\usepackage[a-1b]{pdfx}
\crefname{algocf}{alg.}{algs.}
\Crefname{algocf}{Algorithm}{Algorithms}

\usepackage{amsthm}
\usepackage{multirow}
\usepackage{subfig}
\usepackage{balance}
\newcommand\vldbdoi{10.14778/3717755.3717770}
\newcommand\vldbpages{1118 - 1130}
\newcommand\vldbvolume{18}
\newcommand\vldbissue{4}
\newcommand\vldbyear{2024}
\newcommand\vldbauthors{\authors}
\newcommand\vldbtitle{\shorttitle} 
\newcommand\vldbavailabilityurl{URL_TO_YOUR_ARTIFACTS}
\newtheorem{lemma}{Lemma}
\usepackage{textcomp}
\newcommand{\sik}{SIG-$k$NNG\xspace}
\newcommand{\mbv}{HSIG\xspace}
\newcommand{\nhq}{NHQ\xspace}
\newcommand{\ngt}{NGT\xspace}
\newcommand{\vear}{Vearch\xspace}
\newcommand{\adb}{ADBV\xspace}
\newcommand{\mil}{Milvus\xspace}

\newcommand{\serf}{SeRF\xspace}
\newcommand{\rann}{RF-ANNS\xspace}

\newcommand\vldbpagestyle{empty} 
\newtheorem{definition}{Definition}
\newtheorem{example}{Example}
\newtheorem{theorem}{Theorem}
\newtheorem*{proof*}{Proof}
\usepackage[linesnumbered,ruled,vlined]{algorithm2e}
\usepackage{titlesec}
\titleformat{\subsubsection}
  {\normalfont\normalsize\bfseries}{\thesubsubsection}{1em}{}
\titlespacing*{\subsubsection}
  {0pt}{3.25ex plus 1ex minus .2ex}{1.5ex plus .2ex}
\usepackage{wrapfig}
\renewcommand{\abstractname}{\MakeUppercase{Abstract}}

\begin{document}
\title{UNIFY: Unified Index for Range Filtered Approximate Nearest Neighbors Search} 

\author{Anqi Liang}
\affiliation{%
  \institution{Shanghai Jiao Tong University}
}
\email{lianganqi@sjtu.edu.cn}

\author{Pengcheng Zhang}
\affiliation{%
  \institution{Tencent Inc.}
}
\email{petrizhang@tencent.com}

\author{Bin Yao} 
\authornote{represents the corresponding author.}
\affiliation{%
  \institution{Shanghai Jiao Tong University}
}
\email{yaobin@cs.sjtu.edu.cn}

\author{Zhongpu Chen}
\affiliation{%
  \institution{Southwestern University of Finance and Economics}
}
\email{zpchen@swufe.edu.cn}

\author{Yitong Song}
\affiliation{%
  \institution{Shanghai Jiao Tong University}
}
\email{yitong_song@sjtu.edu.cn}

\author{Guangxu Cheng}
\affiliation{%
  \institution{Tencent Inc.}
}
\email{andrewcheng@tencent.com}






\begin{abstract}
This paper presents an efficient and scalable framework for Range Filtered Approximate Nearest Neighbors Search (\rann) over high-dimensional vectors associated with attribute values. Given a query vector \( q \) and a range \( [l, h] \), \rann aims to find the approximate \( k \) nearest neighbors of \( q \) among data whose attribute values fall within \( [l, h] \). Existing methods including pre-, post-, and hybrid filtering strategies that perform attribute range filtering before, after, or during the ANNS process, all suffer from significant performance degradation when query ranges shift. Though building dedicated indexes for each strategy and selecting the best one based on the query range can address this problem, it leads to index consistency and maintenance issues.

Our framework, called UNIFY, constructs a unified Proximity Graph-based (PG-based) index that seamlessly supports all three strategies. In UNIFY, we introduce SIG, a novel \underline{\textbf{S}}egmented \underline{\textbf{I}}nclusive \underline{\textbf{G}}raph, which segments the dataset by attribute values. It ensures the PG of objects from any segment combinations is a sub-graph of SIG, thereby enabling efficient hybrid filtering by reconstructing and searching a PG from relevant segments. 
Moreover, we present \underline{\textbf{H}}ierarchical \underline{\textbf{S}}egmented \underline{\textbf{I}}nclusive \underline{\textbf{G}}raph (\mbv), a variant of SIG which incorporates a hierarchical structure inspired by HNSW to achieve logarithmic hybrid filtering complexity.
We also implement pre- and post-filtering for \mbv by fusing skip list connections and compressed HNSW edges into the hierarchical graph.
Experimental results show that UNIFY delivers state-of-the-art \rann performance across small, mid, and large query ranges.
\end{abstract}

\maketitle

\pagestyle{\vldbpagestyle}
\begingroup\small\noindent\raggedright\textbf{PVLDB Reference Format:}\\
\vldbauthors. \vldbtitle. PVLDB, \vldbvolume(\vldbissue): \vldbpages, \vldbyear.\\
\href{https://doi.org/\vldbdoi}{doi:\vldbdoi}
\endgroup
\begingroup
\renewcommand\thefootnote{}\footnote{\noindent
This work is licensed under the Creative Commons BY-NC-ND 4.0 International License. Visit \url{https://creativecommons.org/licenses/by-nc-nd/4.0/} to view a copy of this license. For any use beyond those covered by this license, obtain permission by emailing \href{mailto:info@vldb.org}{info@vldb.org}. Copyright is held by the owner/author(s). Publication rights licensed to the VLDB Endowment. \\
\raggedright Proceedings of the VLDB Endowment, Vol. \vldbvolume, No. \vldbissue\ %
ISSN 2150-8097. \\
\href{https://doi.org/\vldbdoi}{doi:\vldbdoi} \\
}\addtocounter{footnote}{-1}\endgroup

\ifdefempty{\vldbavailabilityurl}{https://github.com/sjtu-dbgroup/UNIFY}{
\vspace{.3cm}
\begingroup\small\noindent\raggedright\textbf{PVLDB Artifact Availability:}\\
The source code, data, and/or other artifacts have been made available at \url{https://github.com/sjtu-dbgroup/UNIFY}.
\endgroup
}

\section{INTRODUCTION}
\label{sec:intro}
In recent years, Approximate Nearest Neighbors Search (ANNS) has drawn great attention
for its fundamental role in data mining \cite{valkanas2017mining, mo2022simple}, recommendation systems  \cite{suchal2010full}, and retrieval-augmented generation (RAG) \cite{lewis2020retrieval}, etc.
Numerous ANNS methods \cite{arora2018hd,muja2014scalable,zheng2020pm,nsg,malkov2018efficient,wang2021comprehensive,tian2023db} have been developed to efficiently retrieve similar unstructured objects (e.g., text, images, and videos) by indexing and searching their high-dimensional feature vectors \cite{mo2022simple}.
However, ANNS fails to support many real-world scenarios where users need to filter objects not only by feature similarities but also by certain constraints. For example, Google Image Search allows users to upload an image and search for similar images within a specific period. Likewise, e-commerce platforms such as Amazon enable customers to find visually similar products within a price range.

The above queries can be formulated as the \textit{range filtered approximate nearest neighbors search (\rann)} queries.
Consider a vector dataset where each vector is associated with a numeric attribute (e.g., date, price, or quantity). 
Given a query vector $q$ and a range $[l,h]$, \rann returns $q$'s approximate $k$ nearest neighbors among the data whose attributes are within the range $[l, h]$.
Several studies \cite{wei2020analyticdb, wang2021milvus, wang2024efficient, yang2020pase, qin2020similarity} have been carried out on the \rann problem, which can be categorized into the following three strategies based on when the attribute filtering is performed.

\noindent\textbf{Strategy A: Pre-Filtering.} This strategy performs ANNS after the attribute filtering. For example, Alibaba ADBV \cite{wei2020analyticdb}  integrates the pre-filtering strategy using a B-tree to filter attributes, followed by a linear scan on the raw vectors or PQ \cite{jegou2010product} codes.
Milvus \cite{wang2021milvus} partitions data by attributes and builds ANNS indexes for subsets. \rann is done by first filtering out partitions covering the query range, then performing ANNS on subset indexes. This strategy is efficient for small query ranges, but it does not scale well with larger ranges due to the linearly increasing overhead for scanning qualified vectors or indexes.

\noindent\textbf{Strategy B: Post-Filtering.}
This strategy uses an ANNS index to find $k^\prime$ candidate vectors ($k^\prime > k$), then filters by attributes to obtain the final top-$k$ results. \vear \cite{li2018design} and \ngt \cite{NGTonline} apply this strategy, which can be easily extended to popular ANNS indexes like HNSW \cite{malkov2018efficient} and IVF-PQ \cite{jegou2010product,johnson2019billion}.
This method is efficient for large query ranges. In the extreme case where the query range covers 100\% of objects, \rann becomes equivalent to ANNS, making post-filtering efficient.
However, if the query range is small, the ANNS stage may struggle to collect enough qualified candidates, resulting in sub-optimal performance.

\noindent\textbf{Strategy C: Hybrid Filtering.} 
In Strategy A or B, an \rann query is decomposed into two sub-query systems for attribute filtering and ANNS processing.
In contrast, Strategy C employs a single data structure to index and search vectors and attributes simultaneously.
Several studies \cite{gollapudi2023filtered, wang2024efficient} propose to employ state-of-the-art Proximity Graphs (PGs) \cite{wang2021comprehensive}, e.g., HNSW \cite{malkov2018efficient}  and Vahama \cite{jayaram2019diskann,gollapudi2023filtered}, to implement Strategy C for ANNS with categorical filtration (searching similar vectors whose attributes match a label).
The recent study \serf \cite{zuo2024serf} is the first work implementing Strategy C with PGs for \rann.
It compresses multiple HNSWs into a hybrid index.
\rann is carried out by reconstructing and searching an HNSW index only containing objects in the query range. \serf achieves state-of-the-art performance on query ranges from 0.3\% to 50\%.
However, it lags behind Strategy A and B for smaller and larger ranges due to the overhead of HNSW reconstruction.
Additionally, \serf lacks support for incremental data insertion, limiting its use in scenarios with new object arrivals.

\noindent\textbf{Problems and Our Solutions.} As discussed above, existing methods suffer from two challenges: \textit{sub-optimal performance when query range shifts} and \textit{lack of incremental data insertion support}.
A trivial solution is to build dedicated indexes for each strategy and adaptively select the best one based on the query range. However, this requires extra effort to ensure data consistency across multiple indexes, leading to high maintenance costs \cite{wang2024efficient}.
To tackle these problems, we propose a UNIFY framework combining Strategies A, B, and C into a unified PG-based index supporting incremental insertion.
UNIFY is designed to enhance \rann performance by prioritizing the following key objectives: (O1) enabling efficient hybrid filtering, (O2) supporting incremental index construction, (O3) integrating pre- and post-filtering strategies, and (O4) implementing a range-aware selection of search strategies.
The key techniques in UNIFY to achieve these objectives include:

\textit{(1) Segmented Inclusive Graph (SIG) (O1).} For efficient hybrid filtering, we introduce a novel graph family named SIG. SIG segments the dataset based on attribute values and theoretically guarantees that the PG of objects from any combination of segments is a sub-graph of SIG. For example, segment a dataset $\mathcal{D}$ into three subsets $\mathcal{D}_1$, $\mathcal{D}_2$, and $\mathcal{D}_3$, and let $\mathbb{G}(\mathcal{X})$ denote the PG constructed on dataset $\mathcal{X}$. The PGs $\mathbb{G}(\mathcal{D}_1)$, $\mathbb{G}(\mathcal{D}_2)$, $\mathbb{G}(\mathcal{D}_1 \cup \mathcal{D}_2)$, $\dots$, $\mathbb{G}(\mathcal{D}_1 \cup \mathcal{D}_2 \cup \mathcal{D}_3)$ are all included in SIG.
This characteristic allows us to reconstruct and search a small PG from relevant segments intersected with the query range to enhance \rann performance.

\textit{(2) Hierarchical Segmented Inclusive Graph (\mbv) (O1 \& O2).} Based on SIG, we introduce \mbv, a hierarchical graph inspired by HNSW.
Similar to HNSW, \mbv is built incrementally.
Besides, as a variant of SIG, \mbv can approximately ensure that the HNSW of objects from any combination of segments is a sub-graph of \mbv. By reconstructing and searching an HNSW for relevant segments, we achieve $O(\log(n'))$ \rann time complexity, where $n'$ is the number of objects in those segments.

\textit{(3) Fusion of skip list connections (O3).}
The skip list is a classic attribute index optimized for one-dimensional key-value lookup and range search (see \Cref{sec:preliminaries} for more details). We observe that the skip list shares a similar hierarchical structure with our \mbv.
Inspired by this, we fuse the skip list connections that reflect the order of attribute values into the hierarchical graph structure to form a hybrid index.
By navigating these skip list connections, we can efficiently select objects within the query range, thereby implementing efficient pre-filtering.

\textit{(4) Global edge masking (O3).} Recall that post-filtering relies on a global ANNS index over the entire dataset. \mbv ensures that the global HNSW is approximately a sub-graph of it, meaning that each node's global HNSW edges are included within its \mbv edges. We employ an edge masking algorithm to mark the global HNSW edges with a compact bitmap.
Navigated by these bitmap-marked edges, we can perform ANNS over the global HNSW to efficiently support post-filtering.

\textit{(5) Range-aware search strategy selection (O4).}
Inspired by the effectiveness of Strategies A, B, and C for different query ranges, we developed a heuristic for range-aware strategy selection.
Let $Y$ denote the cardinality of objects that fall within the query range, our heuristic is: 
use Strategy A for $Y \leq \tau_A$, Strategy B for $Y \geq \tau_B$, and Strategy C for $\tau_A <  Y < \tau_B$.
Here, $\tau_A$ and $\tau_B$ serve as thresholds to distinguish the ranges for which each strategy is most effective, and they can be derived from historical data statistics.
In this paper, we run a set of sample queries to collect statistics and determine $\tau_A$ and $\tau_B$.
Experimental results show that this heuristic is effective.

\noindent\textbf{Contributions.} Our contributions are summarized as follows:
\begin{itemize}
\item We introduced SIG, a novel graph family that segments the dataset based on attribute values, ensuring efficient hybrid filtering by allowing the reconstruction and search of a PG from relevant segments.
\item We developed \mbv, a novel hybrid index that supports efficient hybrid filtering with logarithmic time complexity for \rann and enables incremental data insertion.
\item We integrated novel auxiliary structures, including skip list connections and edge masking bitmaps, into \mbv to support both pre- and post-filtering strategies. To the best of our knowledge, \mbv is the first index supporting pre-, post-, and hybrid filtering simultaneously.
\item Experiments on real-world datasets demonstrate that our approach significantly outperforms state-of-the-art methods for query ranges from 0.1\% to 100\% by up to 2.29 times.
\end{itemize}

\section{PRELIMINARIES}\label{sec:preliminaries}
\subsection{Problem Definition}
This paper considers a dataset $\mathcal{D}$ with attributed vectors and nearest neighbors search (NNS) with attribute constraints. Specifically, let $A$ be an attribute (e.g., date, price, or quantity). We use $v[A]$ to denote the attribute value associated with vector $v$. The range filtered nearest neighbors search (RF-NNS) problem is defined as:

\begin{definition}[RF-NNS] Given a dataset $\mathcal{D}$ of $n$ attributed vectors $\{v_1,v_2,\dots,v_n\}$, a distance function $\Gamma(\cdot, \cdot)$, and a query $Q=(q, [l,h], k)$ with $q$ as the query vector, an integer $k$ from 1 to $n$, and $[l, h]$ a real-valued query range, RF-NNS returns the $kNN(q, \mathcal{R})$, a subset of $\mathcal{R}=\{v \mid v \in \mathcal{D}$ and $l \leq v[A] \leq h\}$. For any $o\in kNN(q, \mathcal{R})$ and any $u\in \mathcal{R} \setminus kNN(q,\mathcal{R})$, it holds that $\Gamma(o,q) < \Gamma(u,q)$. If $|\mathcal{R}|<k$, all objects in $\mathcal{R}$ are returned.
\end{definition}


Due to the "curse of dimensionality" \cite{indyk1998approximate}, exact NNS in high-dimensional space is inefficient \cite{li2019approximate}. As a result, most research focuses on approximate nearest neighbors search (ANNS), which reports approximate results with an optimized recall.
Similarly, this paper studies the RF-ANNS problem, which returns an approximate result set $kNN'(q,\mathcal{R})$ with an optimized recall:
\begin{equation}
    recall = \frac{| kNN'(\boldsymbol{q},\mathcal{R}) \cap kNN(\boldsymbol{q},\mathcal{R})|}{\min(k,|\mathcal{R}|)}.
\end{equation}

\begin{algorithm}[!t] \SetKwInOut{Input}{Input}\SetKwInOut{Output}{Output}
\caption{ANNSearch}
\label{alg:anns}
\Input{$\mathbb{G}$: HNSW layer; $q$: query vector; $ep$: entry point; $k$: an integer.} 
      \Output{$q$'s approximate $k$ nearest neighbors on $\mathbb{G}$.}
	 \BlankLine  
        push $ep$ to the min-heap $cand$ in the order of distance to $q$;\\
        push $ep$ to the max-heap $ann$ in the order of distance to $q$;\\
        mark $ep$ as visited;\\
        \While{$|cand| > 0$}{
        {$o\gets$ pop the nearest object to $q$ in $cand$};\\
        {$u\gets$ the furthest object to $q$ in $ann$};\\
        \lIf{$\Gamma(o, q) > \Gamma(u, q)$} { \textbf{break}}
        \ForEach{unvisited $v\in \mathbb{G}[o]$}{
            mark $v$ as visited;\\
            $u\gets$ the furthest object to $q$ in $ann$;\\
            \If{$\Gamma$(v,q) < $\Gamma$(u,q) \text{or} $|ann|$ < k}{
                    push $v$ to $cand$ and $ann$;\\
                    \lIf{$|ann| > k$}{pop $ann$}
            }
        }
        }
        \Return{$ann$}; 
 \end{algorithm}

\subsection{Proximity Graph}
A Proximity Graph (PG) \cite{wang2021comprehensive} treats a vector as a graph node, with connections built based on vector proximity. Various greedy heuristics are proposed to navigate the graph for ANNS \cite{wang2021comprehensive}. In the following, we introduce two PGs related to our work.

\noindent\textbf{$k$ Nearest Neighbor Graph (\textbf{$k$NNG})} \cite{paredes2006practical}. 
Given a dataset $\mathcal{D}$, a $k$NNG is built by connecting each vector $v$ to its $k$ nearest neighbors, $kNN(v, \mathcal{D} \setminus \{v\})$. $k$NNG limits the number of edges per node to at most $k$, making it suitable for memory-constrained environments.
However, $k$NNG focuses on local connections and does not guarantee global connectivity, leading to sub-optimal performance compared to state-of-the-art PGs such as HNSW \cite{malkov2018efficient}.

\begin{algorithm}[t] \SetKwInOut{Input}{Input}\SetKwInOut{Output}{Output}
\Input{$\mathbb{H}$: HNSW; $q$: query vector; $ep$: entry point; $ef$: enlarge factor; $k$: an integer.} 
      \Output{$q$'s approximate $k$ nearest neighbors.}
	 \BlankLine  
        $L\leftarrow$ max level of $\mathbb{H}$;\\

        \ForEach{$ L \geq i \geq 1$}{
            \tcc{Search the $i$-th layer $\mathbb{H}^i$.}
            $ann\leftarrow$ANNSearch($\mathbb{H}^i$, $q$, $ep$, 1);\\
            $ep\leftarrow$the nearest object to $q$ in $ann$;\\
        }
        \tcc{Search the bottom layer $\mathbb{H}^0$.}
        $ann\leftarrow$ANNSearch($\mathbb{H}^0$, $q$, $ep$, $ef$);\\
        \Return{\textnormal{top-}$k$ \textnormal{nearest objects to} $q$ \textnormal{in} $ann$}; 
	 \caption{HierarchicalANNS}
    \label{alg:search}
 \end{algorithm}

\begin{algorithm}[t] \SetKwInOut{Input}{Input}\SetKwInOut{Output}{Output}
\Input{$\mathbb{G}$: HNSW layer; $v$: the object to insert;  $ep$: entry point; $M$: the maximum degree; $efCons$: number of candidate neighbors.} 
      \Output{the updated graph.}
	 \BlankLine  
        $ann \leftarrow$  ANNSearch($\mathbb{G}$, $v$, $ep$, $efCons$) \\
        $\mathbb{G}[v] \leftarrow$  Prune($v$, $ann$, $M$) \\
        \ForEach{$o \in \mathbb{G}[v]$} {
            add $v$ to $\mathbb{G}[o]$; \\
            \lIf{$|\mathbb{G}[o]| > M$}{ $\mathbb{G}[o] \leftarrow$  Prune($o$, $\mathbb{G}[o]$, $M$) }
        }
      \Return{$\mathbb{G}$}; 
	 \caption{InsertLayer}
    \label{alg:hnsw-insert-layer}
 \end{algorithm}

\noindent\textbf{Hierarchical Navigable Small World Graph (HNSW)} \cite{malkov2018efficient}.
HNSW is inspired by the 1D probabilistic structure of the skip list \cite{pugh1990skip}, where each layer is a linked list ordered by 1D values.
The bottom layer includes all objects, while the upper layers contain progressively fewer objects.
HNSW extends this structure by replacing linked lists with PGs, enabling efficient hierarchical search. The hierarchical structure of HNSW is similar to that in \Cref{fig:mbv-hnsw}.
As shown in \Cref{alg:anns}, HNSW uses a greedy approach to find the $k$NN for query vector $q$ in each layer. The neighbors of each node $v_i$ in a given layer $\mathbb{G}$ are stored in adjacency lists $\mathbb{G}[v_i]$. The search starts at an entry point $ep$, the most recently inserted vector at the topmost layer \cite{malkov2018efficient}. It repeatedly selects the nearest object $o$ to $q$ from $cand$ (Line 5), adds $o$'s unvisited neighbors to $cand$ (Lines 8--12), and updates $ann$ (Lines 11--13). The search ends when all nodes in $cand$ are farther from $q$ than those in $ann$.
As shown in \Cref{alg:search}, the hierarchical search in HNSW begins at a coarse layer to identify promising regions and progressively descends to finer layers for detailed exploration. To improve accuracy, \Cref{alg:search} uses the parameter $ef$ ($ef>k$), initially exploring $ef$ neighbors to broaden the search region, and finally returns the top-$k$ results.

HNSW is built incrementally based on \Cref{alg:hnsw-insert-layer}.
When inserting a new object $v$ into an HNSW layer, the process begins by searching for its top $efCons$ nearest neighbors using the \textit{ANNSearch} algorithm (Line 1). The \textit{Prune} heuristic is then applied to limit $v$'s connections to a maximum of $M$ (Line 2). The value of $M$, typically set between 5 and 48 \cite{malkov2018efficient}, balances accuracy and efficiency, with its optimal value determined experimentally. The \textit{Prune} method initially sorts $v$'s candidate neighbors by distance. For each candidate $r$, if a neighbor $e$ satisfies $\Gamma(v,r) > \Gamma(e,r)$, $r$ is pruned. Otherwise, an edge between $v$ and $r$ is inserted. The process continues until $v$ has $M$ neighbors.
Additionally, $v$ is added to the neighbor lists of its identified neighbors, and the \textit{Prune} method is applied to each neighbor to maintain the $M$-neighbor limit (Lines 3--5). The \textit{Prune} strategy prevents the graph from becoming overly dense, ensuring efficient navigation.
HNSW achieves state-of-the-art ANNS time complexity of $O(\log n)$ \cite{wang2021comprehensive} and demonstrates top-tier practical performance indicated by various benchmarks \cite{li2019approximate,aumuller2020ann, wang2021comprehensive}.
It also serves as the backbone for various vector databases such as Pinecone \cite{pinecone}, Weaviate \cite{weaviate}, and Milvus \cite{wang2021milvus}.

\section{SEGMENTED INCLUSIVE GRAPH}\label{sec:MBVoverview}
We aim to develop a PG-based index that integrates three search strategies and supports incremental construction. \Cref{sec:inclusivity} introduces the Segmented Inclusive Graph (SIG), a novel graph family that leverages attribute segmentation and inclusivity for hybrid filtering.
\Cref{sec:sig-knng} explores the challenges of constructing an SIG using the basic PG structure $k$NNG and presents practical solutions.

\subsection{SIG Overview}
\label{sec:inclusivity} 

\noindent\textbf{Attribute Segmentation.} Given a dataset of $n$ objects, \rann is concerned only with the $n'$ objects within the query range, where  $n' \leq n$.
To efficiently filter out the qualified objects, we employ the attribute segmentation method to reduce the search space. Assuming that the attribute distribution remains stable, we first sample objects from the dataset and sort them by their attribute values, then apply an equi-depth histogram \cite{piatetsky1984accurate} to partition them. The histogram bin boundaries define the attribute intervals for each segment, partitioning the dataset into disjoint subsets with similar sizes for further indexing.

\noindent\textbf{Graph Inclusivity and SIG.}
After attribute segmentation, we focus on designing a graph for efficient hybrid filtering. Since a query range intersects a few segments, a straightforward approach is to independently build PGs for each segment. For an \rann query, we search the local PGs in intersected segments and merge the results for the global $k$NN. However, this method results in search time increasing linearly with the number of intersected segments. State-of-the-art PGs offer an ANNS time complexity of $O(\log n)$, suggesting sub-linear search time growth with the number of intersected segments. For example, with $S$ segments, searching multiple PGs scales as $O(S\log(\frac{n}{S}))$, while searching a single PG containing all objects scales as $O(\log n)$. The ratio between them is $S\left(1-\frac{\log S}{\log n}\right)$. Since $1 < S \ll n$, this ratio is greater than 1 but less than $S$, indicating that searching across multiple PGs is less efficient than searching a single PG containing all objects.

Inspired by this, we introduce a novel graph family known as the Segmented Inclusive Graph (SIG). An SIG ensures that the PG of any segment combination is included in (i.e., is a sub-graph of) the SIG. Leveraging this characteristic, termed \textit{graph inclusivity}, allows efficient hybrid filtering by reconstructing and searching the smaller PG of a few segments covering the query range. Below, we introduce the formal definitions of graph inclusivity and SIG.

\begin{definition}[Graph Inclusivity and Segmented Inclusive Graph]
\label{def:sig}
Let $\mathbb{G}(\mathcal{X})$ denote a PG constructed over the dataset $\mathcal{X}$. 
Given a dataset $\mathcal{D}$ segmented into $S$ disjoint subsets $\mathcal{P}$=\{$\mathcal{D}_1, \mathcal{D}_2, \dots, \mathcal{D}_S$\}, a segmented inclusive graph $\mathbb{SIG}(\mathcal{D})$ is a type of graph that possesses the property of \textit{graph inclusivity} defined as

\begin{equation}
\forall r \in \{1,2,\cdots,S\},\ 
\forall \mathcal{C} \in \mathcal{P}^{(r)},\ 
    \mathbb{G}(\bigcup_{e \in \mathcal{C}} e) \subseteq\mathbb{SIG}(\mathcal{D}),
\end{equation}
where $\mathcal{P}^{(r)}$ denotes all possible combinations of $r$ elements from $\mathcal{P}$.

\end{definition}

\begin{example}
Assume that $\mathcal{P}=\{\mathcal{D}_1, \mathcal{D}_2, \mathcal{D}_3\}$, we have $\mathcal{P}^{(2)}=\{\{\mathcal{D}_1,\mathcal{D}_2\}, \{\mathcal{D}_1,\mathcal{D}_3\}, \{\mathcal{D}_2,\mathcal{D}_3\}\}$. For one of the possible combinations $\mathcal{C}$ $=$ $\{\mathcal{D}_1,\mathcal{D}_2\}$, we have $\bigcup_{e \in \mathcal{C}}(e)$ $=$$\mathcal{D}_1 \cup \mathcal{D}_2$.
It holds that $\mathbb{G}(\mathcal{D}_1 \cup \mathcal{D}_2) \subseteq \mathbb{SIG}(\mathcal{D}) $.

\end{example}
Based on \Cref{def:sig}, we derive an exhaustive SIG construction algorithm: build PGs for all segment combinations and merge them into a single graph. While it guarantees graph inclusivity, it faces significant computational challenges. With $S$ segments, the number of combinations is $\sum_{i=1}^{S}\binom{S}{i}=2^S - 1$, and the space complexity of SIG scales as $O(2^S-1)$, making the construction of all PGs both computationally and spatially prohibitive\footnote{In fact, for \rann, we only need to consider continuous segments, with potential combinations totaling $0.5(n^2-n)$, which is also prohibitive.
Since the conclusions in this paper apply to both all segment combinations and continuous ones, we discuss the harder scenario, i.e., all segment combinations, to maintain generality.
}. Thus, we propose a space-efficient SIG index that scales as $O(S)$ in the next section.


\subsection{SIG-$k$NNG}
\label{sec:sig-knng}
\begin{figure}[t]
    \centering
    \includegraphics[width=7cm]{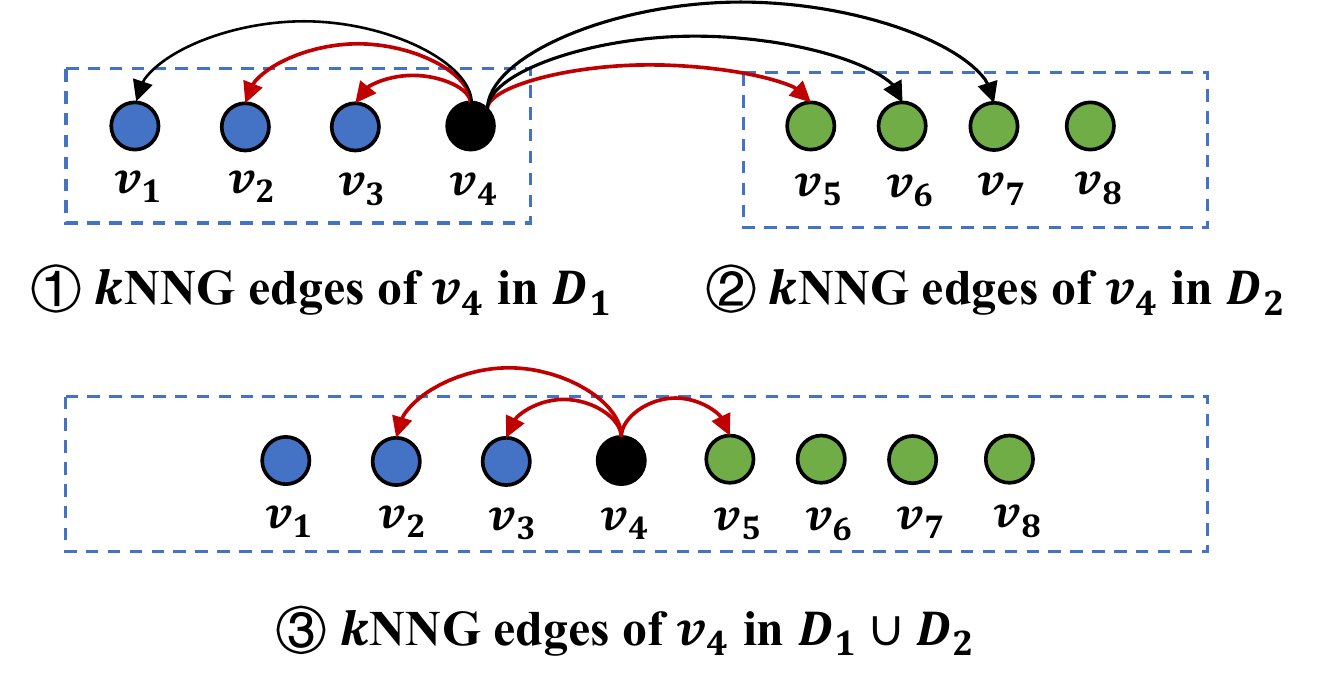}
    \caption{Build multiple $k$NNGs exhaustively $(k=3)$.}
    \label{fig:knng-naive-build}
\end{figure}
\begin{figure}[t]
    \centering
    \includegraphics[width=0.7\linewidth]{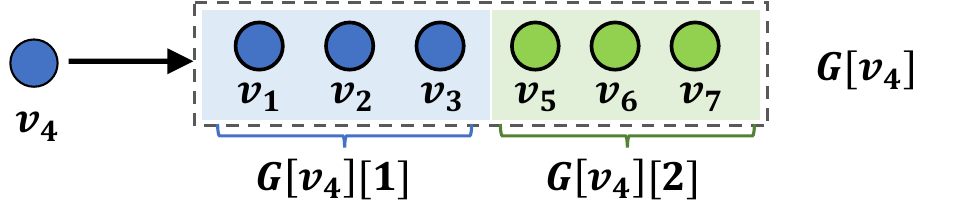}
    \caption{Example of the segmented adjacency list ($k=3$).}
    \label{fig:sig-knng neighbors}
\end{figure}
In this section, we conduct a case study on $k$NNG to demonstrate the limitations of the exhaustive method. Then, we introduce SIG-$k$NNG, a novel graph structure that guarantees inclusivity without exhaustively building all possible $k$NNGs.

\noindent\textbf{Limitation of the Exhaustive Method.}
For each object $v$ in a dataset $\mathcal{D}$, constructing a $k$NNG involves adding a directed edge $(v, o)$ for each object $o \in kNN(v, \mathcal{D} \setminus \{v\})$.
When the dataset is divided into $S$ subsets by attribute segmentation, the exhaustive method requires running the construction algorithm $2^S-1$ times. \Cref{fig:knng-naive-build} illustrates an example.
Here, the dataset $\mathcal{D}$ is divided into two subsets $\mathcal{D}_1=\{v_1, v_2, v_3, v_4\}$ and $\mathcal{D}_2=\{v_5, v_6, v_7, v_8\}$. We need to build $k$NNGs ($k$=3) for $\mathcal{D}_1$, $\mathcal{D}_2$, and $\mathcal{D}_1 \cup \mathcal{D}_2$, respectively. Such exhaustive construction is unnecessary, as the $k$NN of a union is inherently within the individual $k$NN sets.
For example, all objects in $kNN(v_4, \mathcal{D}_1 \cup \mathcal{D}_2)$ can be found in $kNN(v_4, \mathcal{D}_1)$ $\cup$ $kNN(v_4, \mathcal{D}_2)$.
In other words, $v_4$'s three edges in graph \textcircled{3} are already included in graphs \textcircled{1} and \textcircled{2}, making the construction of graph \textcircled{3} redundant.

\noindent\textbf{Structure of \sik.}
Inspired by the prior observations, we introduce \sik, a novel graph structure that achieves inclusivity without the need for exhaustive $k$NNG construction. 
\sik uses a segmented adjacency list to store the outgoing edges for each object in the graph. As illustrated in \Cref{fig:sig-knng neighbors}, given a dataset $\mathcal{D}$ segmented into $S$ subsets $\mathcal{D}_1, \dots, \mathcal{D}_S$, the adjacency list for any object $v$ is divided into $S$ chunks. For example, the full adjacency list $\mathbb{G}[v_4]$ of object $v_4$ is segmented into two chunks $\mathbb{G}[v_4][1]$ and $\mathbb{G}[v_4][2]$. The $i$-th chunk stores only $v$'s $k$NN within $\mathcal{D}_i$. Based on this design, to add SIG-$k$NNG edges for $v$, we perform $k$NN searches $S$ times to find its neighbors in each subset, instead of searching $2^S-1$ times across all subset combinations.
In the following, we introduce the formal definition of \sik and prove that \sik can exactly guarantee inclusivity.

\begin{definition}[\sik] Given a dataset $\mathcal{D}$ segmented into $S$ disjoint subsets $\mathcal{D}_1, \mathcal{D}_2, \dots, \mathcal{D}_S$,
the \sik on $\mathcal{D}$ is defined as $\mathbb{F}_k(\mathcal{D}) = (V_\mathbb{F}, E_\mathbb{F})$, where $V_\mathbb{F} = \mathcal{D}$ and $E_\mathbb{F} =\bigcup_{v\in\mathcal{D}} \bigcup_{i=1}^{S} E_{i}^{k}(v)$. Here, $E_{i}^{k}(v)$ is the set of edges based on the $i$-th chunk of $v$'s adjacency list, defined as $E_{i}^{k}(v) = \{(v, o) \mid o \in kNN(v, \mathcal{D}_i \setminus \{v\})\}$.
\label{def:mbv-knng}
\end{definition}

\noindent\textbf{Inclusivity of \sik.}
We first present the following lemma, which shows that the $k$NN of a union is inherently included in the individual $k$NN sets.
\begin{lemma}
    \label{theorem:knn-union}
    Given $r$ disjoint datasets $\mathcal{D}_1, \mathcal{D}_2, \dots, \mathcal{D}_r$ and their union set $\mathcal{U} = \bigcup_{i=1}^r{\mathcal{D}_i}$, it holds that
     $\forall v \in \mathcal{U}$,  $k$NN$(v, \mathcal{U}\setminus\{v\}) \subseteq $
     $\bigcup_{i=1}^{r}{kNN(v, \mathcal{D}_{i} \setminus \{v\})}$.
\end{lemma}

\begin{figure}[t]
    \centering
    \includegraphics[width=\linewidth]{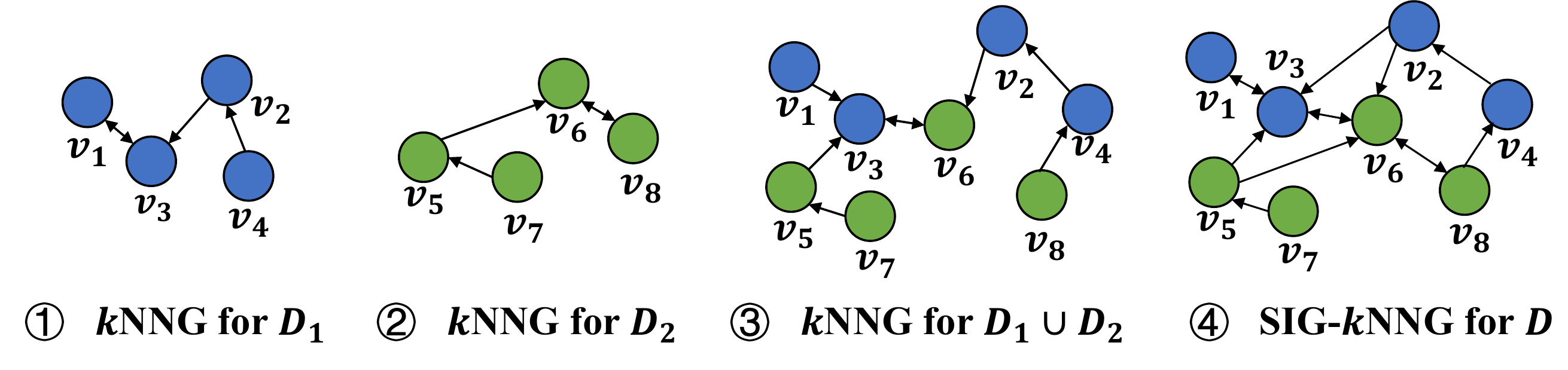}
    \caption{Illustration of SIG-$k$NNG's inlusivity ($k=1$).}
    \label{fig:sigknng-demo}
\end{figure}
Based on \Cref{theorem:knn-union}, we derive the inclusivity of \sik, as presented in \Cref{theorem:knn-inclusivity}. The proof is omitted here for brevity, as it is straightforward to follow.
\begin{theorem}
\label{theorem:knn-inclusivity}
Let $\mathbb{G}_k(\mathcal{X})$ and $\mathbb{F}_k(\mathcal{X})$ denote a $k$NNG and \sik for dataset $\mathcal{X}$, respectively. Given a dataset $\mathcal{D}$ segmented into $S$ disjoint subsets $\mathcal{D}_1, \mathcal{D}_2, \dots, \mathcal{D}_S$, it follows that \sik is a segmented inclusive graph.
Specifically, for any $r$ distinct integers $i_1, i_2, \ldots, i_r$ chosen from $[1, S]$ with $1 \leq r \leq S$, we have $\mathbb{G}_k(\mathcal{U}) \subseteq \mathbb{F}_k(\mathcal{D})$, where $\mathcal{U}=\bigcup_{j=1}^{r} \mathcal{D}_{i_j}$.
\end{theorem}

\begin{example}
\Cref{fig:sigknng-demo} demonstrates the inclusivity of \sik. Given a dataset $\mathcal{D} = \{v_1, v_2, \dots, v_8\}$, we assume the attribute of $v_i$ is $i$ for simplicity. The attribute space $[1, 8]$ is divided into two disjoint segments $[1, 5)$ and $[5, 8]$, leading to two subsets $\mathcal{D}_1=\{v_1, v_2, v_3, v_4\}$ and $\mathcal{D}_2=\{v_5, v_6, v_7, v_8\}$.
Graphs \textcircled{1}\textasciitilde\textcircled{3} represent the $k$NNGs for $\mathcal{D}_1$, $\mathcal{D}_2$, and $\mathcal{D}_1 \cup \mathcal{D}_2$, and the \sik for $\mathcal{D}$ is displayed in graph \textcircled{4}. As illustrated, all three $k$NNGs are sub-graphs of the \sik.
\end{example}

\begin{figure}[!t]
  \centering
  \includegraphics[width=\linewidth]{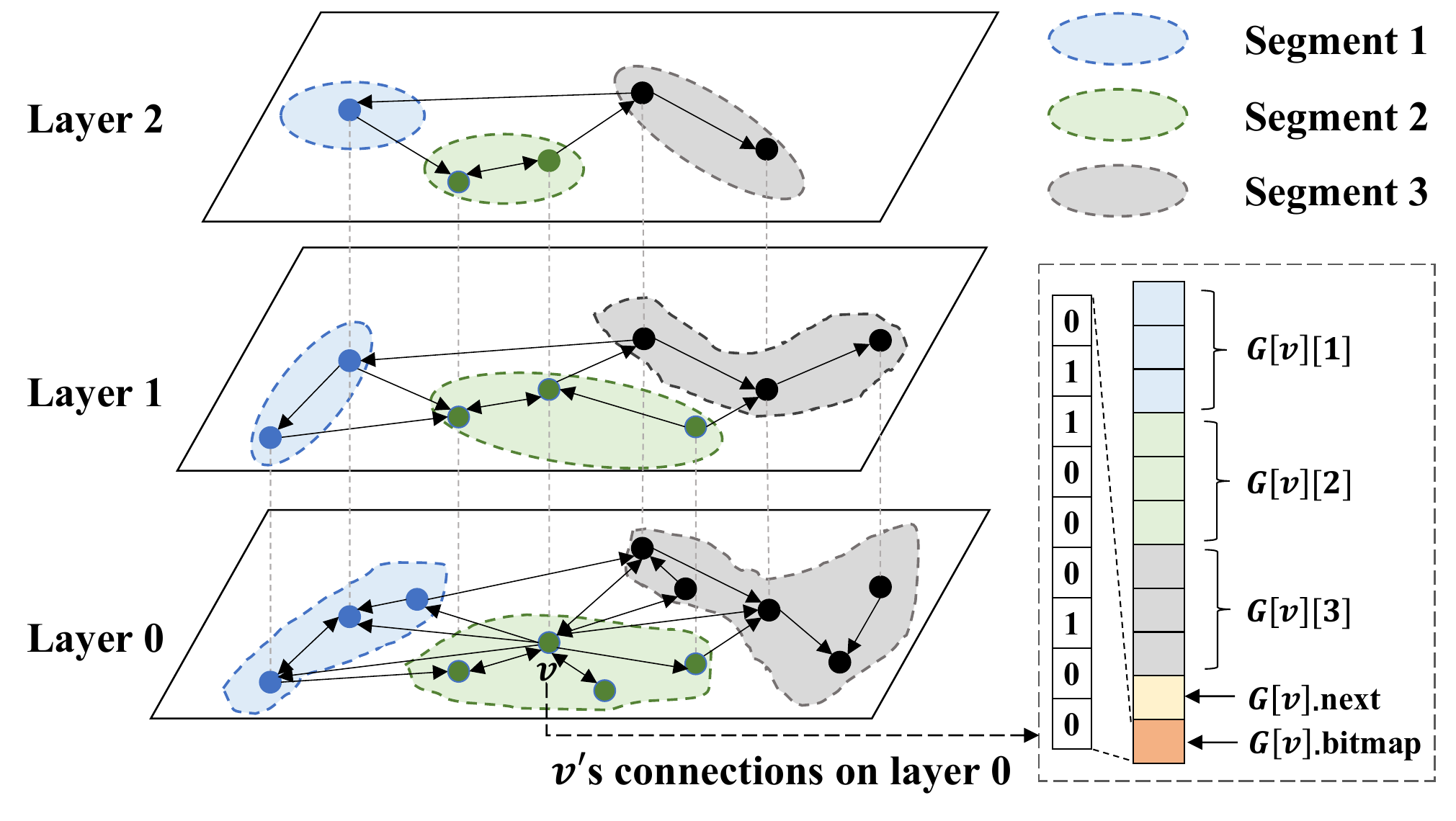}
  \caption{Illustration of \mbv.}
  \label{fig:mbv-hnsw}
\end{figure}

\section{HIERARCHICAL SEGMENTED INCLUSIVE GRAPH} \label{sec:hsig}
\sik offers an efficient approach to build an SIG with segmented adjacency lists. Though it theoretically guarantees inclusivity, the underlying $k$NNG is not competitive with state-of-the-art PGs like HNSW \cite{malkov2018efficient}.
Additionally, \sik lacks support for incremental insertion. In this section, we expand the basic idea of \sik, introducing the Hierarchical Segmented Inclusive Graph (\mbv), which uses HNSW as a building block to achieve incremental construction and a logarithmic search complexity.

\Cref{fig:mbv-hnsw} provides an overview of \mbv, showcasing a hierarchical structure that follows the structure of HNSW and skip lists. \mbv is a unified structure that indexes both vectors and attributes by leveraging the strengths of vector-oriented HNSW and attribute-oriented skip lists. For vector indexing, \mbv organizes the outgoing edges of each object into chunks based on attribute segmentation, inspired by \sik. Each chunk contains the edges of an HNSW for the corresponding segment. Thus, the backbone graph of \mbv can be seen as a set of HNSWs constructed for different segments (e.g., the three HNSWs with different colors in \Cref{fig:mbv-hnsw}), which are mutually connected with additional edges to ensure inclusivity. Additionally, we incorporate the skip list into the backbone graph to index attributes for efficient pre-filtering and introduce a compact auxiliary structure to optimize post-filtering. The structure and algorithms are detailed in \Cref{sec:hsig cons,sec:hsig-search}.

\subsection{\mbv Construction}\label{sec:hsig cons}
In this section, we describe the core structures of HSIG for hybrid, pre-, and post-filtering sequentially. Then, we integrate them to introduce the complete insertion algorithm.

\noindent\textbf{Backbone Graph for Hybrid Filtering.}
Based on \Cref{theorem:knn-inclusivity}, we present \mbv's backbone graph, which is designed to ensure inclusivity regarding HNSW and support efficient hybrid filtering.
As depicted in \Cref{alg:hnsw-insert-layer},
when inserting a new object $v$ into an HNSW layer, the core operation involves finding $v$'s $efCons$ nearest neighbors and establishing up to $M$ connections.
This procedure is similar to $k$NNG's construction, which finds $k$NN and establishes at most $k$ connections. 
Therefore, we use the segmented adjacency list introduced by \sik to guarantee the inclusivity for all possible HNSW connections. With $S$ segments, connections are stored in $S$ chunks, with a maximum degree of $M$ per chunk. The values of $S$ and $M$ are experimentally determined, with default values set to 8 and 16, respectively.
Since each chunk contains the HNSW connections in the corresponding segment, the objects and connections in chunk $j$ form a sub-graph $\mathbb{G}_j$ of the backbone graph in \mbv. When inserting object $v$ that is located in segment $i$, \Cref{alg:insert1} outlines how to build $v$'s connections in segment $j$ at an \mbv layer.
First, we search for $v$'s $efCons$ nearest neighbors in $\mathbb{G}_j$ (Line 1) using \textit{ANNSearch} (\Cref{alg:anns}), then select up to $M$ neighbors using the \textit{Prune} algorithm. For each neighbor $o$, we establish mutual connections between $v$ and $o$ within their respective segments (Lines 2--4). If the number of neighbors of $o$ in segment $i$ exceeds $M$, we apply the \textit{Prune} algorithm to discard the extra neighbors (Lines 5--6). This algorithm guarantees the approximate inclusivity of \mbv, as demonstrated in \Cref{sec:inclusive}.

\begin{algorithm}[!t]
\SetKwInOut{Input}{Input}
\SetKwInOut{Output}{Output}
\Input{$\mathbb{G}$: \mbv layer; $ep$: entry point; $v$: the object to insert; $i$: index of the segment that $v[A]$ belongs to; $j$: index of the segment to insert; $M$: the maxinum degree; $efCons$: number of candidate neighbors.}
\Output{$v$'s approximate nearest neighbors in $\mathbb{G}_j$ (a sub-graph of $\mathbb{G}$ with only nodes and edges in segment $j$).}

$ann  \leftarrow$ ANNSearch($\mathbb{G}_j$, $v$, $ep$, $efCons$); \\
\ForEach{$o \in$ \textnormal{Prune}$(v, ann, M) $} {
    add $(v, o)$ to $\mathbb{G}[v][j] $;\\ 
    add $(o, v)$ to $\mathbb{G}[o][i] $;\\
    \If{$\mid \mathbb{G}[o][i]\mid > M$} {
        $\mathbb{G}[o][i] \leftarrow $ Prune($o$, $\mathbb{G}[o][i]$, $M$); \\
    }
}
\Return $ann$;
\caption{BackboneConnectionsBuild}
\label{alg:insert1}
\end{algorithm}

\noindent\textbf{{Fusing Skip List Connections for Pre-Filtering.}} For small-range \rann queries, pre-filtering with an attribute index typically outperforms PG-based methods.
This is because PGs prioritize vectors over attributes, making it difficult to filter candidates within the query range. For instance, in the extreme case of a query range containing only one object, using an attribute index to quickly locate the target object is optimal. Thus, we propose fusing an attribute index into the graph structure for effective pre-filtering. 
As described in \Cref{sec:preliminaries}, the skip list is a popular index for efficient 1D key-value lookups and range searches, sharing a hierarchical structure similar to HNSW and \mbv. 
Inspired by this, we integrate the skip list into the backbone graph of \mbv to form a unified index.
Navigated by skip list connections, we can efficiently locate and linearly search objects within the query range. As shown in \Cref{fig:mbv-hnsw}, we use an extra connection $\mathbb{G}[v].next$ to store the ID of $v$'s successor object in the skip list. After inserting an object into the backbone graph via \Cref{alg:insert1}, we search for its successor in the skip list and store it in $\mathbb{G}[v].next$.

\noindent\textbf{Global Edge Masking for Post-Filtering.} For large-range queries, post-filtering with a pure ANNS index is preferable. We use a global HNSW over the entire dataset to support post-filtering. Due to inclusivity, the global HNSW is (approximately) a sub-graph of \mbv. We apply a global edge pruning method to identify these HNSW edges from \mbv edges and use a compact bitmap to mask the unused edges. Since each segment's sub-graph has a maximum degree of $M$, we similarly limit the connections in the global graph to a maximum of $M$. 
This procedure is detailed in \Cref{alg:insert3}. First, we obtain the object $v$'s connections in all segments and select $M$ connections using the \textit{Prune} algorithm (Line 1). Next, we create a bitmap whose size equals the number of $v$'s neighbors in all segments and set the positions of $v$'s $M$ neighbors to 1 (Line 2). For each neighbor $o$ in the $M$ neighbors (Line 3), if $v$ is also a neighbor of $o$, set the position of $v$ in $o$'s bitmap to 1 (Lines 4--6). Then, we update the bitmap of $o$ (Lines 7--9).


\begin{algorithm}[t]
\SetKwInOut{Input}{Input}
\SetKwInOut{Output}{Output}
\Input{$\mathbb{G}$: \mbv layer; $v$: the object to insert; $M$: the maximum degree.}
\Output{the updated graph.}
\BlankLine

$\mathcal{N} \leftarrow$ Prune($v$, $\mathbb{G}[v]$, $M$);\\
$\mathbb{G}[v].bitmap \leftarrow$ generate bitmap with $\mathcal{N}$;\\
\ForEach{$o \in \mathcal{N}$} {
    \If{$v \in \mathbb{G}[o]$}{
        $pos \leftarrow$ $v$'s position in $\mathbb{G}[o]$;\\
        $\mathbb{G}[o].bitmap[pos] \leftarrow 1$;\\
        \If{ $sum(\mathbb{G}[o].bitmap) > M$} {
          $\mathcal{N}^\prime \leftarrow$ Prune($o$, $\mathbb{G}[o]$, $M$);\\
          $\mathbb{G}[o].bitmap \leftarrow$ update bitmap with $\mathcal{N}^\prime$;\\
        }
    }
}
\Return $\mathbb{G}$;
\caption{GlobalEdgeMasking}
\label{alg:insert3}
\end{algorithm}

\noindent\textbf{Complete Insertion Algorithm.} As aforementioned, the adjacency list of a node in \mbv consists of (1) several chunks for backbone graph connections, (2) a skip list connection, and (3) a bitmap for global connections.
The bottom-right of \Cref{fig:mbv-hnsw} illustrates an example. Here, $\mathbb{G}[v][i]$ is the backbone graph connections of $v$ in the $i$-{th} segment, $\mathbb{G}[v]$.next stores the ID of $v$'s successor in the skip list, and $\mathbb{G}[v]$.bitmap stores the global edge masks. 

Next, we present the complete insertion algorithm for constructing an \mbv. The \mbv structure consists of a set of HNSWs for different segments, augmented with auxiliary structures for pre- and post-filtering. Construction involves three main steps: building the HNSW in each segment, adding skip list connections, and applying global edge masking. Each step is performed hierarchically and incrementally, as outlined in \Cref{alg:insert4}.
First, we find the segment to which object $v$ belongs (Line 1). The maximum layer $level$ of $v$ is randomly assigned using an exponentially decaying probability distribution normalized by $m_L$ (Line 2; see \cite{malkov2018efficient} for details). We then traverse all segments (Line 3) to sequentially build the HNSW $\mathbb{HG}_j$ in each segment $j$, determining its maximum level $L$ and entry point $ep$ (Lines 4--5). For each layer $l$ from $L$ to $level+1$, we use \textit{ANNSearch} (\Cref{alg:anns}) to find the nearest neighbor $ep$ of $v$ in $\mathbb{HG}_j^l$, where $\mathbb{HG}_j^l$ denotes layer $l$ in $\mathbb{HG}_j$ (Lines 6--7). 
Then, for each layer $l$ from $level$ to 0, we employ \textit{BackboneConnectionsBuild} (\Cref{alg:insert1}) to insert $v$ into $\mathbb{HG}_j^l$, using its nearest neighbor as the entry point for the next layer (Lines 8--10). After building $\mathbb{HG}_j$, we update its entry point if necessary (Line 11), ensuring the first object in the topmost layer serves as the entry. Next, $v$ is inserted into the skip list using the method in \cite{pugh1990skip} (Line 12). Finally, we use \textit{GlobalEdgeMasking} (\Cref{alg:insert3}) to mask unnecessary edges for post-filtering from layer 0 to $level$ (Lines 13--14). 

\begin{algorithm}[t]
\SetKwInOut{Input}{Input}
\SetKwInOut{Output}{Output}
\Input{$\mathbb{HG}$: \mbv; $v$: the object to insert; $S$: number of segments; $M$: the maximum degree; $efCons$: number of candidate neighbors.}
\Output{the updated \mbv.}
\BlankLine
$i \leftarrow$ ComputeSegmentId($v[A]$); \\
$level \leftarrow$ $\lfloor -\ln(\mathit{unif}(0 \dots 1)\cdot m_L) \rfloor$;\\
\ForEach{$1 \leq j \leq S $} {
    $L\leftarrow$ the max level of $\mathbb{HG}_j$;\\
    $ep\leftarrow$ the entry point of $\mathbb{HG}_j$;\\
    \ForEach{$ L \geq l \geq level+1$}{
    $ep\leftarrow$ ANNSearch($\mathbb{HG}^l_j$, $v$, $ep$, 1);\\
    }
    \ForEach{$level \geq l \geq 0$}{
        $ann \leftarrow$ BackboneConnectionsBuild($\mathbb{HG}^l_j$, $ep$, $v$, $i$, $j$, $M$, $efCons$);\\
        $ep \leftarrow$ nearest object to $v$ in $ann$;\\
    }
    update graph entry for $\mathbb{HG}_j$ if necessary;\\
}
add skip list connections for $v$ in layers $0\dots level$; \\
\ForEach{$0 \leq l \leq level$ } {
$\mathbb{HG}^l \leftarrow$ GlobalEdgeMasking($\mathbb{HG}^l $, $v$, $M$);\\
}
\Return $\mathbb{HG}$;
\caption{HSIGInsert}
\label{alg:insert4}
\end{algorithm}

\subsection{Search on \mbv}\label{sec:hsig-search}
To adapt to different query ranges, we propose three search strategies and a range-aware search strategy selection method.

\noindent\textbf{Strategy A (Pre-Filtering)}.
Navigated by the hierarchical skip list connections, we can quickly reach the bottom layer to identify the first object whose attribute value falls within the query range. From there, a linear search is performed to collect the query vector's $k$NN among those vectors with qualified attribute values. This method is straightforward, so pseudocode is omitted.

\noindent\textbf{Strategy B (Post-Filtering)}. 
We employ the hierarchical search scheme in \Cref{alg:search} for ANNS, followed by filtering objects within the query range. ANNS is performed over a global HNSW with edges marked by bitmaps. During the search, only outgoing edges marked as 1 in an object's bitmap are considered. The search retrieves the top-$ef$ nearest neighbors ($ef > k$) in the bottom layer, after which attribute filtering is applied to obtain the top-$k$ results. 

\noindent\textbf{Strategy C (Hybrid Filtering)}.
We perform hybrid filtering by reconstructing and searching the HNSW of segments covering the query range.
The search follows the hierarchical scheme in \Cref{alg:search}, starting from the topmost entry point of graphs in all segments. Since each object has $M$ connections per segment, and assuming there are $S'$ segments intersecting the query range, this requires visiting $MS'$ neighboring nodes, leading to a large search space and high computational complexity.
Thus, we introduce the search parameter $m$ to reconstruct an HNSW with a maximum degree of $m$ at runtime. We propose two neighbor selection strategies to select $m$ neighbors from $MS'$ connections: (1) compute distances between node $v$ and its neighbors across all $S^\prime$ segments, then select the top-$m$ neighbors by sorting them by distance; and (2) select the top-($\lceil m/{ S^\prime \rceil})$ neighbors from each chunk of the adjacency list. Since neighbors in each chunk are naturally ordered by distance to $v$ upon acquisition via \textit{ANNSearch}, we simply take the first $\lceil m/{ S^\prime \rceil}$ objects. We adopt the second strategy as the default, based on experimental results in \Cref{sec:neighbor-select}. \Cref{alg:hybridfilter} presents the pseudocode for hybrid filtering at a specific layer. Compared with \Cref{alg:anns}, the key differences are: (1) only examining outgoing edges within intersected segments (Line 8) and (2) pushing qualified objects in the query range into the results (Line 15).

\noindent\textbf{Range-aware Strategy Selection}.
Let $Y$ be the cardinality of objects within a query range.
Observing that pre-, post-, and hybrid filtering perform best for small-, large-, and mid-range queries, respectively, we propose the following heuristic for strategy selection: use Strategy A if $Y \leq \tau_A$, Strategy B if $Y \geq \tau_B$, and Strategy C if $\tau_A < Y < \tau_B$. 
Here, $\tau_A$ and $\tau_B$ are thresholds distinguishing the optimal ranges for each strategy, derived from historical data analysis.
Given these thresholds, we estimate the cardinality of an incoming query to apply the heuristic.
Since statistic collection and cardinality estimation are well studied \cite{harmouch2017cardinality,oommen2002efficiency} and not the focus of this paper, we provide a simple preprocessing method to validate our heuristic. Specifically, we sample objects from the base dataset as queries and assign each a random query range.
We then execute these queries using the three strategies and record recall and latency metrics. Given a recall target, we analyze records meeting this requirement and identify two turning points where pre- and post-filtering outperform hybrid filtering. 
These points establish $\tau_A$ and $\tau_B$, guiding strategy selection for future queries.


\begin{algorithm}[t]
\SetKwInOut{Input}{Input}
\SetKwInOut{Output}{Output}
\Input{$\mathbb{G}$: \mbv layer; $q$: query vector; $[l,h]$: query range; $m$: number of visited neighbors per object; $ep$: entry point; $k$: number of nearest neighbors.}
\Output{$q$'s approximate $k$ nearest neighbors within $[l,h]$.}
push $ep$ to the min-heap $cand$ in the order of distance to $q$;\\
push $ep$ to the max-heap $ann$ in the order of distance to $q$;\\
$\mathcal{S} \leftarrow$ segments that intersect with $[l,h]$;\\
\While{$|cand| > 0$}{
{$o\gets$ pop the nearest object to $q$ in $cand$};\\
{$u\gets$ the furthest object to $q$ in $ann$ };\\
\lIf{$\Gamma(o, q) > \Gamma(u, q)$} { \textbf{break} }
\ForEach{$i \in \mathcal{S}$} {
    $\mathcal{N} \leftarrow$  the top-($\lceil m/|\mathcal{S}| \rceil$) neighbors in $\mathbb{G}[o][i]$;\\
    \ForEach{unvisited $v \in \mathcal{N}$} {
        mark $v$ as visited;\\
            $u\gets$ the furthest object to $q$ in $ann$;\\
            \If{$\Gamma$(v,q) < $\Gamma$(u,q) \text{or} $|ann|$ < k}{
                    push $v$ to $cand$;\\
                    \lIf{$l\leq v[A]\leq h$} { push $v$ to $ann$}
                    \lIf{$|ann| > k$}{pop $ann$ }
            }
        }
}
}
\Return{$ann$}; 
\caption{HybridFilteringLayer}
\label{alg:hybridfilter}
\end{algorithm}

\subsection{Theoretical Analysis}\label{sec:hsig-analysis}
\noindent\textbf{Space Complexity.} Following HNSW \cite{malkov2018efficient}, we use a 32-bit integer to store each edge in \mbv and analyze space complexity using 32-bit as a storage unit. In \mbv, each node has up to $MS$ edges (taking $MS$ units), a bitmap of size $MS$ (taking $\frac{MS}{32}$ units), and a skip list connection (taking one unit).
For a dataset of $n$ objects, \mbv contains $nL'$ nodes, where $L'$ is the average number of levels. This results in an expected space complexity of $O(nL'(\frac{33}{32}MS+1))$. As discussed in \cite{malkov2018efficient}, the average number of levels in HNSW is a constant, and since \mbv uses the same strategy to determine the number of levels, $L'$ is also a constant in \mbv. $S$ and $M$ are experimentally determined and are generally small constants. Given that $L'$, $S$, and $M$ are all considered small constants relative to $n$, the space complexity can be simplified to $O(n)$.

\noindent\textbf{Construction Complexity}. 
Consider a dataset with $n$ objects divided into $S$ subsets, each containing $n_1, n_2, \dots, n_S$ objects. Inserting an object into \mbv requires three operations: (OP1) backbone graph insertion, (OP2) skip list insertion, and (OP3) edge masking. OP3 runs in constant time, while OP2 has an expected complexity of $O(\log n)$ \cite{pugh1990skip}. OP1 performs ANNS on the HNSW of each segment, requiring $S$ ANNS iterations. Since ANNS on an HNSW with $x$ objects takes $O(\log x)$ time, OP1 costs $\sum_{i=1}^S\log(n_i)$ bounded by $S\log n$,  making OP1 an $O(S\log n)$ operation. Neglecting constants, the expected complexity of inserting an object is $O(\log n)$, and constructing an \mbv for $n$ objects scales as $O(n\log n)$.

\noindent\textbf{Search Complexity}.
The complexity scaling of a single search can be strictly analyzed under the assumption that \mbv exactly guarantees inclusivity with respect to HNSW. For an \mbv with $n$ objects, consider a query range covering $Y$ objects and a small constant $k$ negligible compared to $n$. The search complexities of the three strategies in \mbv are as follows:

\textit{A. Pre-Filtering.} Searching the skip list has an expected complexity of $O(\log n)$ \cite{pugh1990skip}, and computing vector distances within the query range takes $O(Y)$ time. Thus, the total complexity is $O(Y+\log n)$. 
Due to range-aware search in \mbv, Pre-Filtering is used only when $Y$ is smaller than a constant $\tau_A$, ensuring it typically operates with an $O(\log n)$ complexity.

\textit{B. Post-Filtering.} This strategy involves searching the global HNSW, which has an expected time complexity of $O(\log n)$ \cite{malkov2018efficient,wang2021comprehensive}.

\textit{C. Hybrid Filtering.} For \rann, hybrid filtering selects segments covering the query range and performs ANNS on their HNSWs. Let $n'$ be the number of objects in the intersected segments, searching the HNSW takes $O(\log n')$ time. Since $n' \leq n$, the expected time complexity is $O(\log n)$. Although the theoretical complexity is derived under the exact inclusivity assumption, experimental results (\Cref{sec:exp-scale}) confirm the method's logarithmic scaling with data size, validating its efficiency and scalability.

\begin{table}[!t] 
        \centering
        \caption{Dataset specifications}
        \label{tab:hybrid dataset}
        \setlength{\tabcolsep}{2pt}
        \begin{tabular}{lccccc}
        \toprule
        Dataset & Dimension & \#Base & \#Query & Type \\
        \midrule
        SIFT1M & 128 & 1,000,000 & 1,000 & Image + Attributes\\
        GIST1M & 960 &1,000,000 & 1,000 & Image + Attributes\\
        GloVe & 100 & 1,183,514 & 1,000 & Text + Attributes\\
        Msong & 420 & 992,272 & 200 &Audio + Attributes\\
        WIT-Image & 2048 & 1,000,000 & 1,000 &Image + Attributes\\
        Paper & 200 & 2,029,997 & 10,000 &Text + Attributes\\
        \bottomrule
        \end{tabular}
\end{table}

\section{EXPERIMENT}\label{sec:exp}
\subsection{Experimental Setup}

\textbf{Datasets.}
We use six real-world datasets of varying sizes and dimensions. The Paper \cite{wang2024efficient} and WIT-Image \cite{zuo2024serf} datasets include both feature vectors and attributes. The Paper contains publication, topic, and affiliation attributes, which we convert from categorical to numerical, while WIT-Image uses image size as its attribute. For the remaining datasets, which contain only feature vectors, we generate numerical attributes using a method similar to \cite{wang2021milvus}, assigning each vector a random value between 0 and 10,000. The dataset characteristics are detailed in \Cref{tab:hybrid dataset}. For each query, we generate a query range uniformly between 0.1\% and 100\%.

\begin{table*}
  \centering
\caption{Index build time and index size.}
\label{tab:indexres}
\begin{tabular}{c|cccccc|cccccc}
\hline
\multirow{2}{*}{Method} & \multicolumn{6}{c|}{Build Time (s)}                                                                                                                      & \multicolumn{6}{c}{Index Size (MB)}                                                                                                                      \\ \cline{2-13} 
                        & \multicolumn{1}{c|}{SIFT1M}       & \multicolumn{1}{c|}{GIST1M}          & \multicolumn{1}{c|}{GloVe}        & \multicolumn{1}{c|}{Msong}        & 
                        \multicolumn{1}{c|}{WIT-Image}       & Paper        & \multicolumn{1}{c|}{SIFT1M} & \multicolumn{1}{c|}{GIST1M} & \multicolumn{1}{c|}{GloVe} & \multicolumn{1}{c|}{Msong}& 
                        \multicolumn{1}{c|}{WIT-Image}       & Paper \\ \hline
\vear    & \multicolumn{1}{c|}{\textbf{735}}          & \multicolumn{1}{c|}{\textbf{1319}} & \multicolumn{1}{c|}{1187}                 & \multicolumn{1}{c|}{2241}  & \multicolumn{1}{c|}{1294}    & \textbf{617}    & \multicolumn{1}{c|}{692}    & \multicolumn{1}{c|}{3905}   & \multicolumn{1}{c|}{741}   & \multicolumn{1}{c|}{2095} & \multicolumn{1}{c|}{4456} & 2430   \\ \hline
\ngt     & \multicolumn{1}{c|}{789}          & \multicolumn{1}{c|}{27357}          & \multicolumn{1}{c|}{15281}        &  \multicolumn{1}{c|}{\textbf{771}}& \multicolumn{1}{c|}{8620} & 814 & \multicolumn{1}{c|}{764}    & \multicolumn{1}{c|}{4031}   & \multicolumn{1}{c|}{773}   & \multicolumn{1}{c|}{1894}& \multicolumn{1}{c|}{4277} &2129   \\ \hline
\nhq     & \multicolumn{1}{c|}{2806}         & \multicolumn{1}{c|}{1689}& \multicolumn{1}{c|}{4956} & \multicolumn{1}{c|}{841} & \multicolumn{1}{c|}{\textbf{889}} &5039& \multicolumn{1}{c|}{78}& \multicolumn{1}{c|}{66}& \multicolumn{1}{c|}{52}    &\multicolumn{1}{c|}{99} & \multicolumn{1}{c|}{97} &158   \\ \hline
\adb     & \multicolumn{1}{c|}{860}          & \multicolumn{1}{c|}{4318}          & \multicolumn{1}{c|}{\textbf{896}} & \multicolumn{1}{c|}{2069} & \multicolumn{1}{c|}{10039}& 2444     & \multicolumn{1}{c|}{\textbf{21}}     & \multicolumn{1}{c|}{\textbf{24}}      & \multicolumn{1}{c|}{\textbf{25}}      & \multicolumn{1}{c|}{\textbf{22}} & \multicolumn{1}{c|}{\textbf{24}}  & \textbf{43} \\ \hline
\mil     & \multicolumn{1}{c|}{1459}         & \multicolumn{1}{c|}{6931}          & \multicolumn{1}{c|}{1560}         &  \multicolumn{1}{c|}{4289} & \multicolumn{1}{c|}{4983}&2477        & \multicolumn{1}{c|}{30}     & \multicolumn{1}{c|}{52}     & \multicolumn{1}{c|}{35}    & \multicolumn{1}{c|}{36}& \multicolumn{1}{c|}{56} &61    \\ \hline
\serf     & \multicolumn{1}{c|}{2502}         & \multicolumn{1}{c|}{11820}                & \multicolumn{1}{c|}{2678}        & \multicolumn{1}{c|}{4817} & \multicolumn{1}{c|}{13440} &6189        & \multicolumn{1}{c|}{763}   & \multicolumn{1}{c|}{3896}      & \multicolumn{1}{c|}{704}     & \multicolumn{1}{c|}{1852} & \multicolumn{1}{c|}{4185}  &2096   \\ \hline
\mbv     & \multicolumn{1}{c|}{2406}         & \multicolumn{1}{c|}{10827}         & \multicolumn{1}{c|}{2601}        & \multicolumn{1}{c|}{4230} &\multicolumn{1}{c|}{16076} &6254         & \multicolumn{1}{c|}{1554}   & \multicolumn{1}{c|}{4728}      & \multicolumn{1}{c|}{1713}     & \multicolumn{1}{c|}{2647} & \multicolumn{1}{c|}{5008} & 3785   \\ \hline
\end{tabular}
\end{table*}

\begin{figure*}[!t]
  \centering
\includegraphics[width=\linewidth]{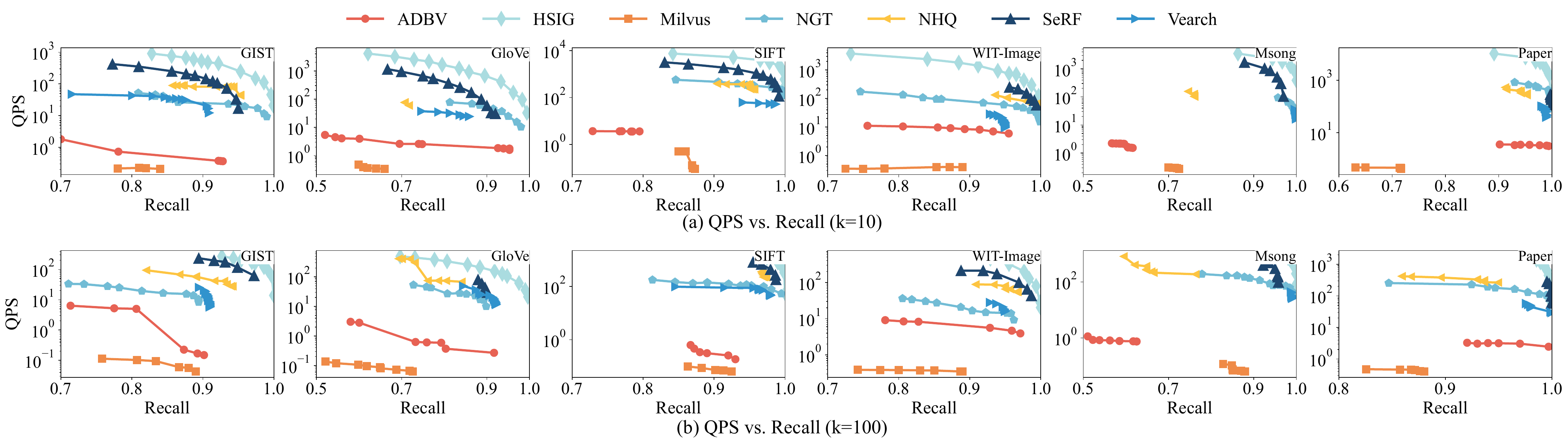}
  \caption{Overall Performance.}
  \label{fig:overall}
\end{figure*}

\noindent\textbf{Compared Methods.} We compare \mbv against six competitors in terms of \rann performance:
\begin{itemize}
\item\textbf{\adb} \cite{wei2020analyticdb} is a hybrid analytic engine developed by Alibaba. It enhances PQ \cite{jegou2010product} for hybrid ANNS and proposes the accuracy-aware, cost-based optimization to generate optimal execution plans.
\item\textbf{\mil} \cite{wang2021milvus} partitions datasets based on commonly utilized attributes and implements \adb within each subset.
\item\textbf{\nhq} \cite{wang2024efficient} constructs a composite graph index based on the fusion distance of vectors and attributes for hybrid queries. It proposes enhanced edge selection and routing mechanisms to boost query performance.
\item\textbf{\ngt} \cite{NGTonline} is an ANNS library developed by Yahoo Japan that processes hybrid queries using the post-filtering strategy.
\item\textbf{\vear} \cite{li2018design,vearchcode} is a high-dimensional vector retrieval system developed by Jingdong that supports hybrid queries through the post-filtering strategy. 
\item\textbf{\serf} \cite{zuo2024serf} designs a 2D segment graph that compresses multiple ANNS indexes for half-bounded range queries and extends this to support general range queries.
\end{itemize}
We use the Euclidean distance function to measure vector distances.

\noindent\textbf{Metrics.}
We evaluate query effectiveness by recall and efficiency by measuring the number of queries processed per second (QPS).

\noindent\textbf{Parameter Settings.} The parameters $S$, $M$, and $efCons$ represent the number of segments, maximum edge connections, and candidate neighbors during index construction, respectively. We use grid search to determine their optimal values, setting $S$, $M$, and $efCons$ to 8, 16, and 500, respectively. The parameters $m$ and $ef$ relate to the search process, where $m$ is the total number of neighbors visited per object, and $ef$ is the number of candidates searched during a query. We vary $m$ and $ef$ to generate recall/QPS curves and apply grid search to set baseline parameters.

\noindent\textbf{Implementation Settings.} We implement \mbv construction and search algorithms based on hnswlib \cite{malkov2018efficient}. The code is written in C++ and compiled with GCC 10.3.1 using the "-O3" optimization flag. A Python interface is provided for the indexing library, and experiments are conducted using Python 3.8.17.  

\noindent\textbf{Environment.} Scalability experiments are conducted on Alibaba Cloud Linux 3.2104 LTS with 40 cores and 512GB memory. Other experiments are conducted on a Linux server with an Intel(R) Xeon(R) E5-2609 v3 (1.90GHz, 6 cores), 16GB memory, and Ubuntu 18.04.5.

\subsection{Overall Performance} \label{sec:exp-overall}
We evaluate the query performance of \mbv and its competitors with $k$ values of 10 and 100. Based on the preprocessing method in \Cref{sec:hsig-search}, we set $\tau_A$ and $\tau_B$ to 1\% and 50\% of the dataset size.
\Cref{fig:overall} shows query performance, while \Cref{tab:indexres} presents index sizes and build times. Although \mbv does not excel in index size or build time due to its multi-segment structure and extensive edge connections, it is comparable to \serf, the state-of-the-art PG-based solution for \rann. PQ-based methods like \adb and \mil are space-efficient but struggle with query accuracy and efficiency compared to PG-based methods.
As shown in \Cref{fig:overall}, \mbv consistently outperforms baselines across all datasets regarding the QPS vs. recall trade-off. For example, with $k=10$ and a recall of around 0.9, \mbv achieves a QPS two orders of magnitude higher than \adb on the GloVe and GIST1M datasets, one order of magnitude higher than \nhq on the SIFT1M dataset, and outperforms \serf by up to 2.29 times across all datasets.
These results highlight \mbv's effectiveness, benefiting from its unified graph structure and range-aware strategy selection. 
Additionally, \mbv allows HNSW reconstruction with varying edge degrees using the parameter $m$, whereas \serf reconstructs HNSW with a fixed edge degree, limiting its query performance.  
Finally, \mbv outperforms \ngt, \vear, and \nhq, as they employ the post-filtering strategy and perform poorly on small query ranges. 

\subsection{Effect of Range-aware Search Strategy Selection} \label{sec:exp-range}
We evaluate \mbv's performance across small, medium, and large query ranges. The methods compared are as follows:
(1) \textbf{\mbv-pre} uses \mbv with the pre-filtering strategy. (2) \textbf{PQ-pre} performs attribute filtering first, followed by PQ-based vector retrieval. (3) \textbf{Btree-pre} uses a B-tree for attribute filtering, followed by brute-force vector retrieval. (4) \textbf{\mbv-post} applies \mbv with the post-filtering strategy. (5) \textbf{HNSW-post} builds HNSW over the entire dataset and applies the post-filtering strategy. (6) \textbf{\mbv-hybrid} employs \mbv for \rann queries using the hybrid filtering strategy. (7) \textbf{\serf} is a state-of-the-art \rann solution that uses a hybrid filtering. (8) \textbf{\mbv-range-aware} uses \mbv with range-aware search strategy selection. (9) \textbf{Dedicated} builds specialized indexes for each strategy, with Btree-pre, HNSW-post, and \serf used for pre-, post-, and hybrid filtering, respectively, and selects the best strategy based on query range.

\begin{figure}[!t]
    \centering
    \begin{minipage}[t]{1.0\linewidth}
        \centering
        \includegraphics[width=0.8\linewidth]{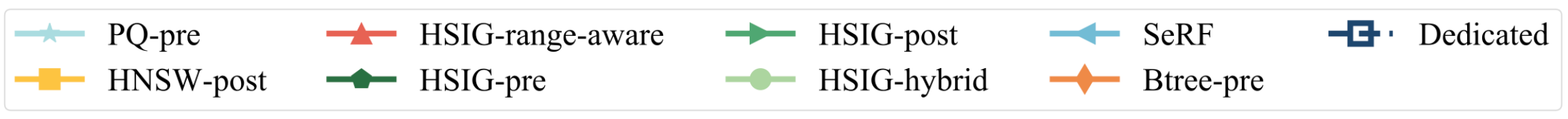}  
    \end{minipage}

    
    {
        \label{fig:select1}{{\selectfont}}
        \begin{minipage}[t]{0.34\linewidth}
            \centering
 \includegraphics[width=\linewidth]{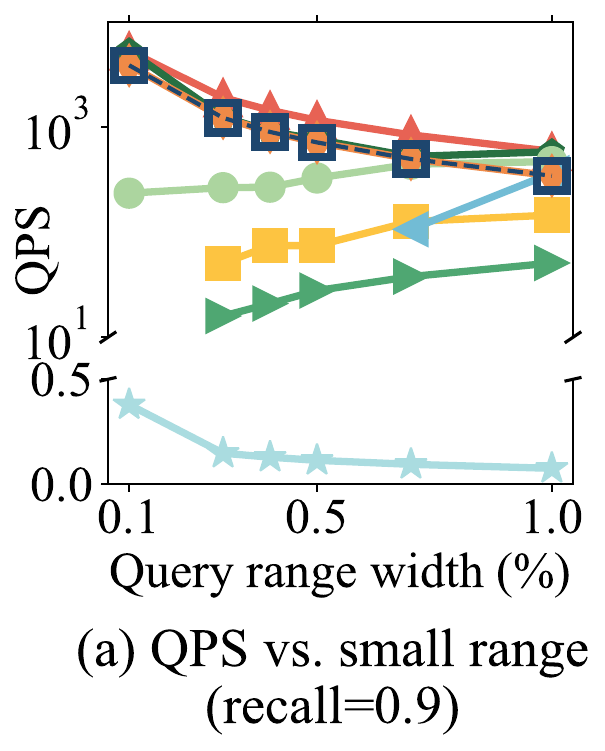}
        \end{minipage}
    }
    {   \label{fig:select2}
    \hspace{-0.46cm}
        \begin{minipage}[t]{0.33\linewidth}
            \centering
            \includegraphics[width=\linewidth]{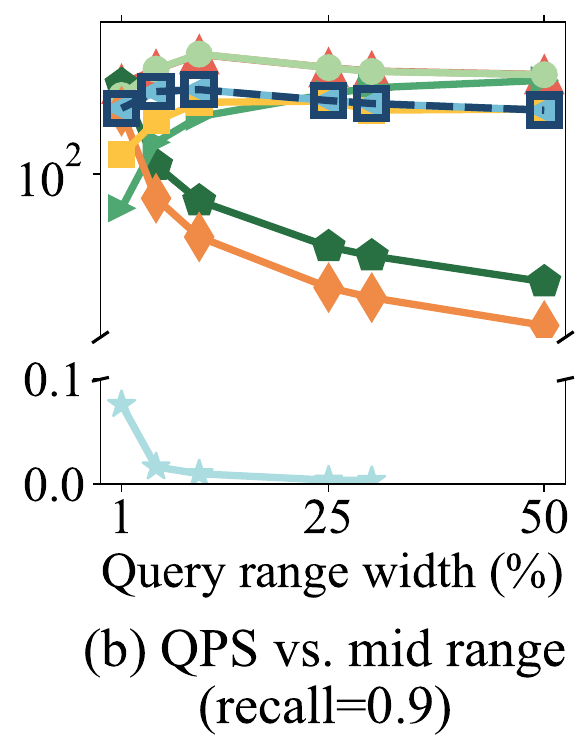}
        \end{minipage}
    }
    {   \label{fig:select2}
    \hspace{-0.46cm}
        \begin{minipage}[t]{0.34\linewidth}
            \centering
            \includegraphics[width=\linewidth]{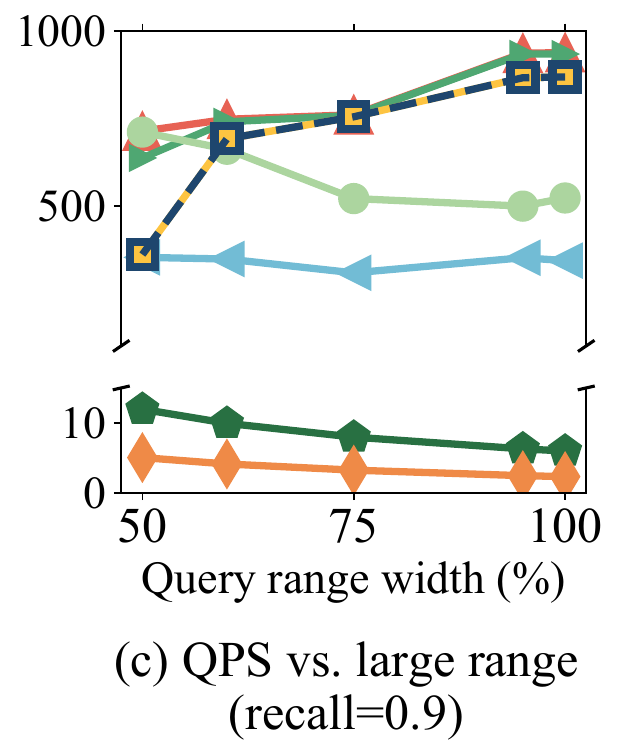}
        \end{minipage}
    }
\vspace{-0.2cm}

    {
        \label{fig:select1}{{\selectfont}}
        \begin{minipage}[t]{0.34\linewidth}
            \centering
            \includegraphics[width=\linewidth]{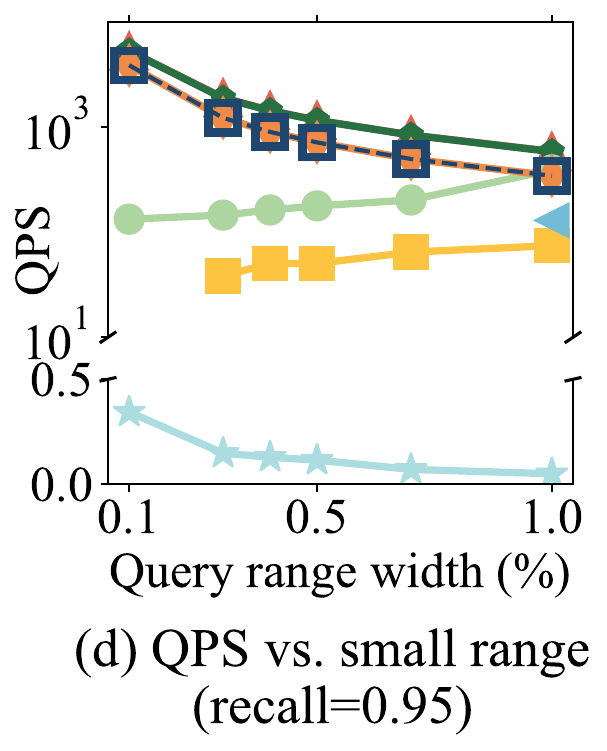}
        \end{minipage}
    }
  {
        \label{fig:select2}
        \hspace{-0.46cm}
        \begin{minipage}[t]{0.34\linewidth}
            \centering
            \includegraphics[width=\linewidth]{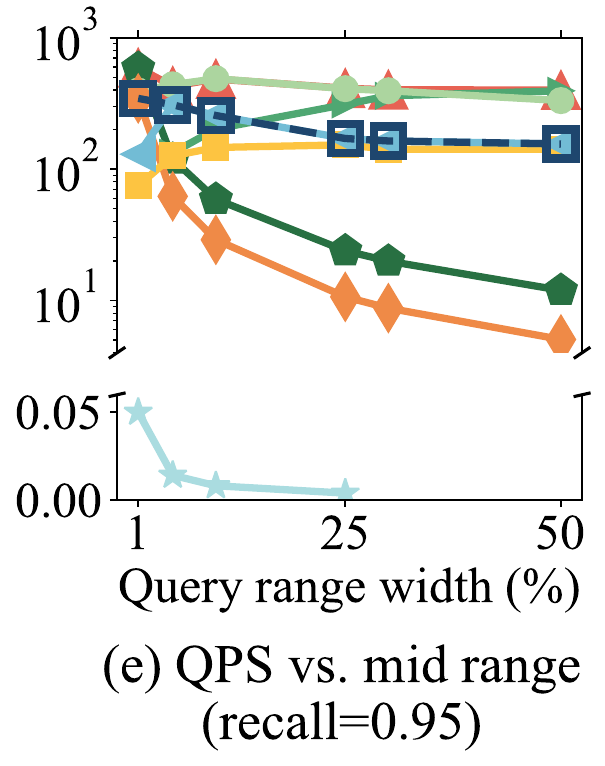}
        \end{minipage}
    } 
    {   \label{fig:select2}
    \hspace{-0.46cm}
        \begin{minipage}[t]{0.34\linewidth}
            \centering
            \includegraphics[width=\linewidth]{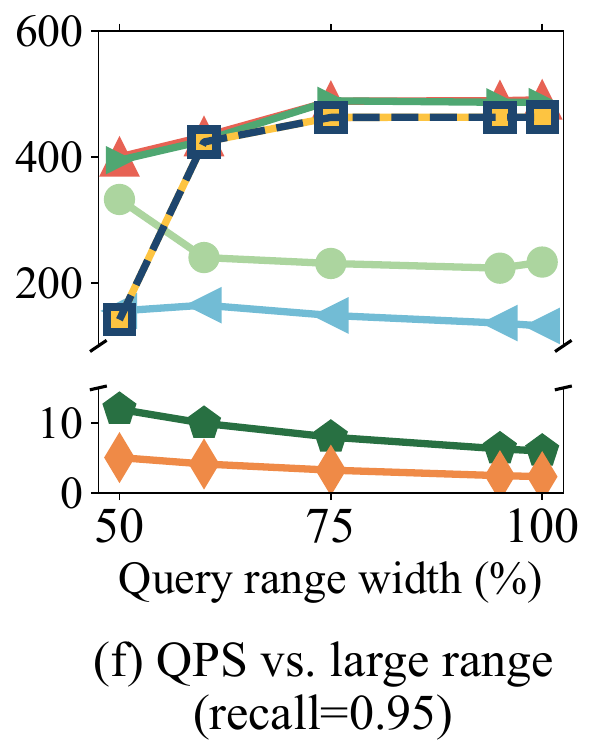}
        \end{minipage}
    }
    \vspace{-0.2cm}
    
    {
        \label{fig:efqps}
        \begin{minipage}[t]{0.34\linewidth}
            \centering
            \includegraphics[width=\linewidth]{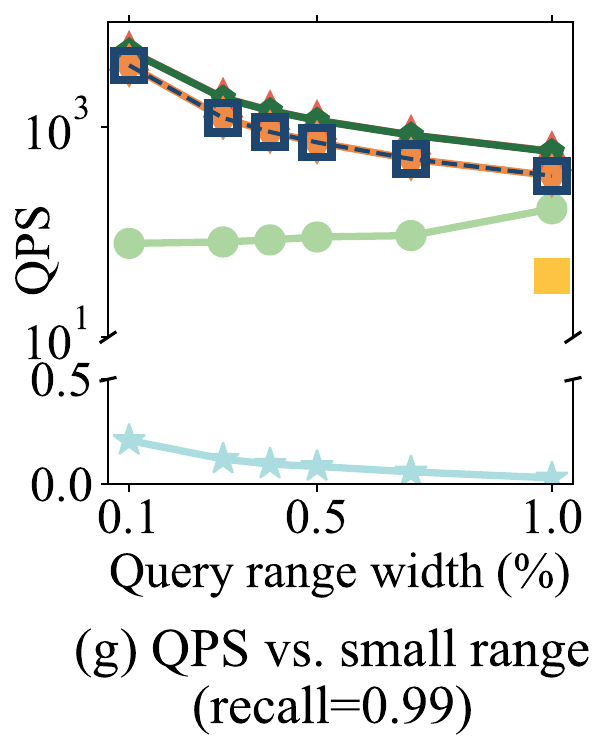}
        \end{minipage}
    }
   {
        \label{fig:alqps}
        \hspace{-0.46cm}
        \begin{minipage}[t]{0.34\linewidth}
            \centering
            \includegraphics[width=\linewidth]{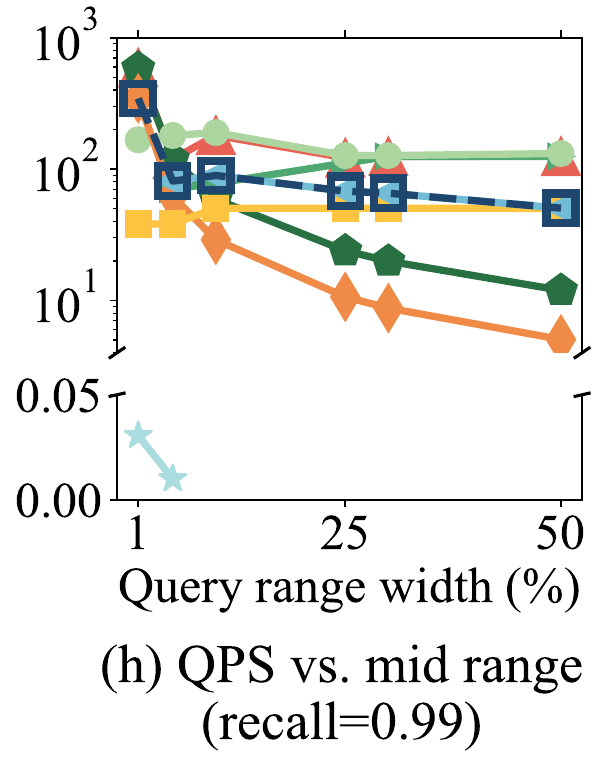}
        \end{minipage}
    }
    {   \label{fig:select2}
    \hspace{-0.46cm}
        \begin{minipage}[t]{0.34\linewidth}
            \centering
            \includegraphics[width=\linewidth]{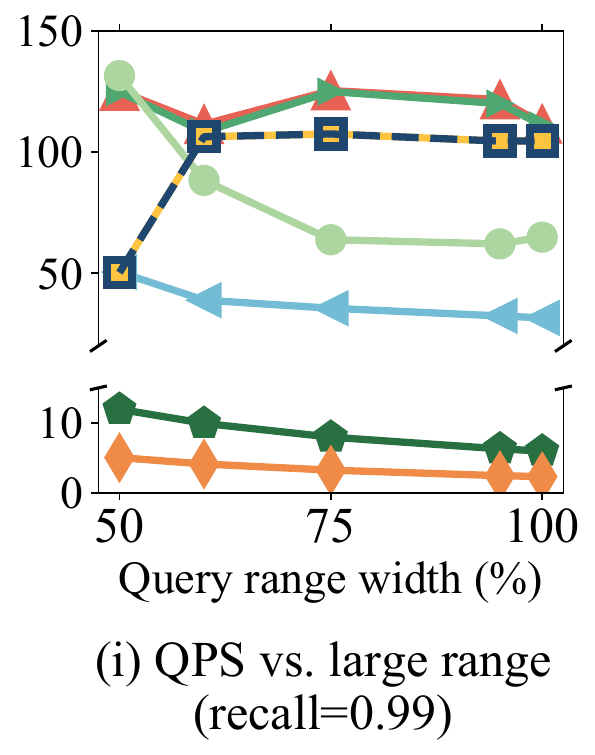}
        \end{minipage}
    }
    \caption{Impact of different query ranges on GloVe dataset.}
    \label{fig:queryrange}
\end{figure}

The results are shown in \Cref{fig:queryrange}. Some methods have missing QPS values for specific query ranges, indicating they could not meet the recall threshold. \mbv-pre, \mbv-hybrid, and \mbv-post outperform competitors in small, medium, and large ranges, respectively. For example, \mbv-pre outperforms Btree-pre by 20.3\% at a recall of 0.9 in small query ranges. \mbv-hybrid exceeds \serf by 1.21 times in medium query ranges at a recall of 0.95. \mbv-post surpasses HNSW-post by 37.5\% at a recall of 0.99 in large query ranges. Additionally, \mbv-pre outperforms \mbv-hybrid and \mbv-post in small ranges, \mbv-post surpasses \mbv-hybrid and \mbv-pre in large ranges, and \mbv-hybrid performs best among the three strategies in medium ranges, validating our range-aware heuristics. Finally, \mbv-range-aware consistently outperforms Dedicated by up to 1.1 times across all query ranges, thanks to its unified PG-based index and range-aware strategy selection.

\subsection{Validation of Inclusivity of \mbv}\label{sec:inclusive}
\begin{figure}[!t]
  \centering
  \includegraphics[width=\linewidth]{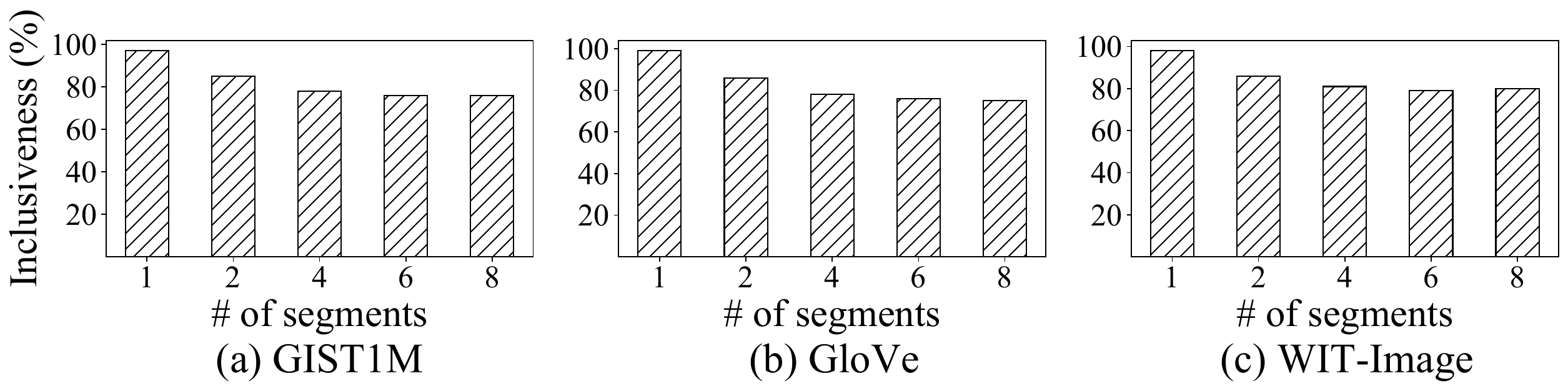}
  \caption{Inclusiveness of \mbv.}
  \label{fig:inclusiveness}
\end{figure}

We evaluate \mbv's inclusivity using the inclusiveness metric. Inclusiveness is computed as $\frac{\#\text{common-edge}}{\#\text{hnsw-edge}}\times 100\%$, where $\#\text{common-edge}$ is the number of identical edge connections in both \mbv and the multi-segment HNSW (where multi-segment HNSW refers to the HNSW constructed for any combination of segments), and $\#\text{hnsw-edge}$ is the total number of edges in the multi-segment HNSW. According to \Cref{def:sig}, the multi-segment HNSW should be a sub-graph of \mbv to satisfy inclusivity. Thus, 100\% inclusiveness indicates that \mbv strictly satisfies inclusivity. In this experiment, we partition the dataset evenly into eight segments and construct HNSWs for 1, 2, 4, 6, and 8 contiguous segments. \Cref{fig:inclusiveness} shows that the average inclusiveness of \mbv exceeds 80\%, demonstrating that \mbv achieves significant inclusiveness and approximately satisfies inclusivity.

\begin{figure}[!t]
  \centering
  \includegraphics[width=\linewidth]{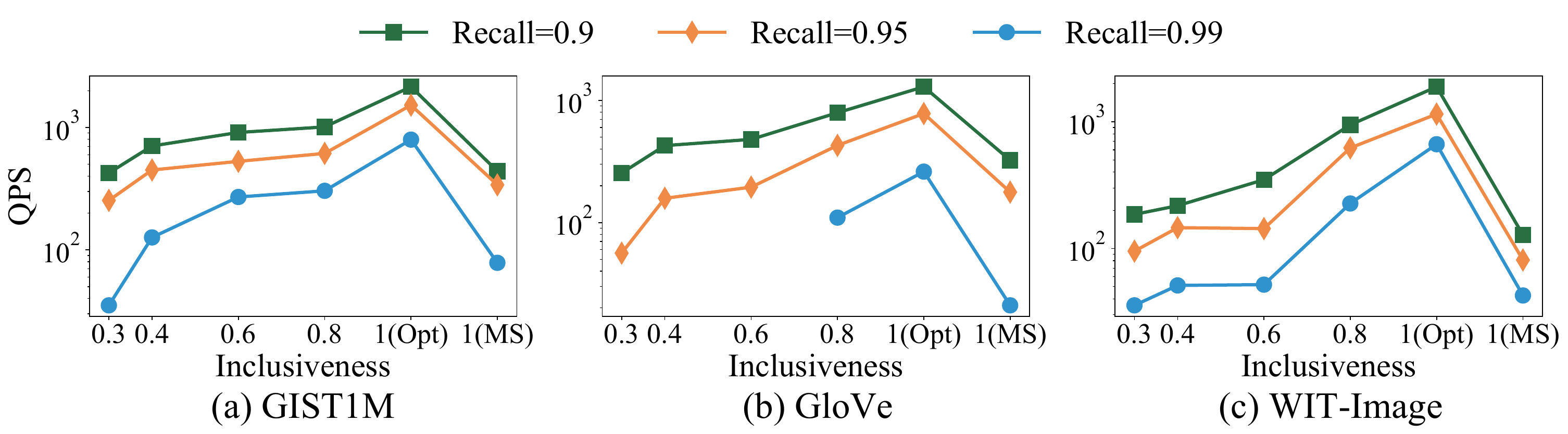}
  \caption{Impact of Inclusivity.}
  \label{fig:inclusivequery}
\end{figure}

To further evaluate the impact of inclusivity on query performance, we compare \mbv at varying levels of inclusiveness (30\%, 40\%, 60\%, and 80\%) against two competitive methods that guarantee exact inclusivity. The first method, \textit{Optimal HNSW}, builds an HNSW in real time for objects within each query range. The second method, \textit{MS-HNSW}, pre-builds HNSWs for each segment. During the search, \textit{MS-HNSW} identifies the segments intersecting with the query range, retrieves vectors from the corresponding HNSWs, and combines the intermediate results to obtain the final results. The results are shown in \Cref{fig:inclusivequery}. While \textit{Optimal HNSW} offers the best performance, building indexes in real time for every query is time-consuming and impractical. As \mbv's inclusiveness increases, its query performance improves, approaching that of \textit{Optimal HNSW}. Additionally, \mbv consistently outperforms \textit{MS-HNSW}, which requires more distance computations. These results demonstrate the effectiveness of \mbv's inclusivity, showing that higher inclusiveness leads to better query performance.

\begin{figure}[!t]
  \centering
\includegraphics[width=0.8\linewidth]{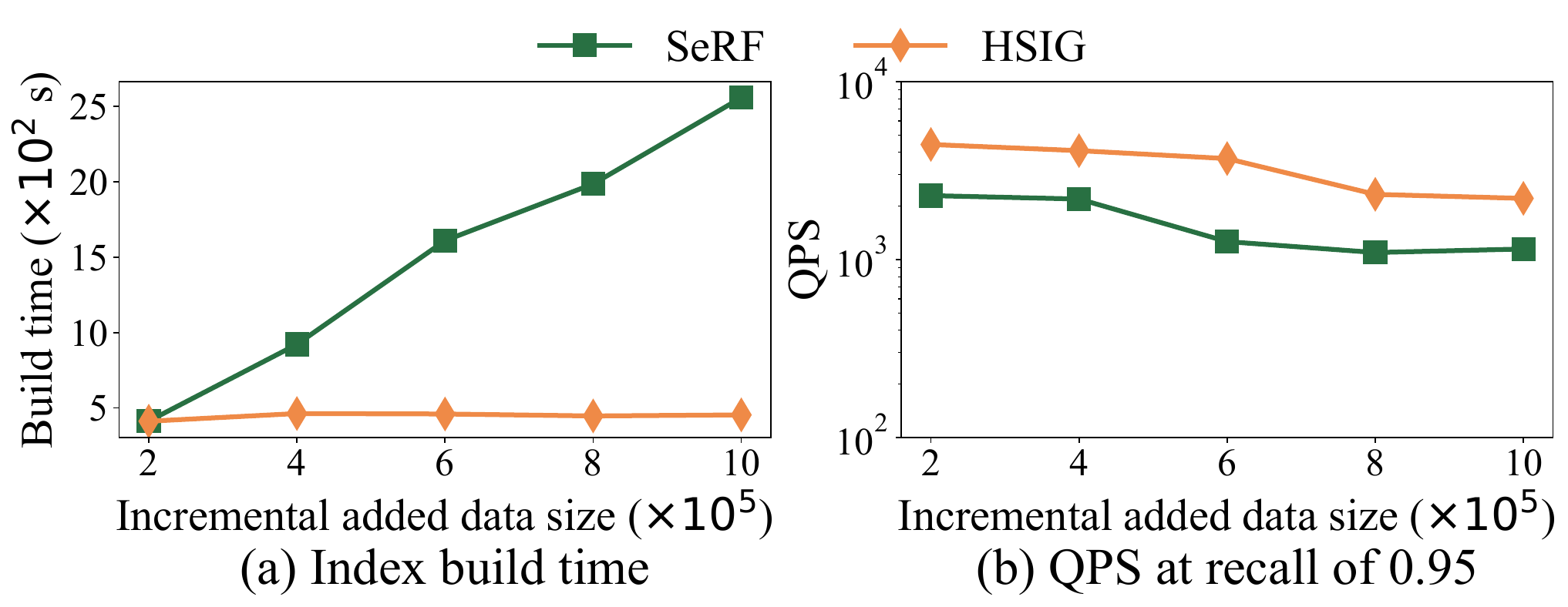}
  \caption{Performance of incremental insertion on SIFT1M.}
  \label{fig:increment}
\end{figure}

\subsection{Validation of Incremental Insertion}
\label{sec:incre-insert}
In this section, we evaluate \mbv's incremental insertion capability. We first build \mbv with 200,000 objects from the SIFT1M dataset, followed by four rounds of incremental insertion, each adding 200,000 objects. We compare the index build time and query performance against the state-of-the-art method \serf. As shown in \Cref{fig:increment}, \mbv's build time remains stable with each insertion since the number of inserted objects is consistent, whereas \serf's build time increases linearly due to its lack of incremental update support. Moreover, \mbv outperforms \serf in query efficiency at a recall of 0.95. These results highlight \mbv's effectiveness in supporting incremental insertions, showing that it is suitable for applications with continuously evolving data.

\subsection{Validation of Runtime Neighbor Selection}\label{sec:neighbor-select}
\begin{figure}[!t]
  \centering
\includegraphics[width=\linewidth]{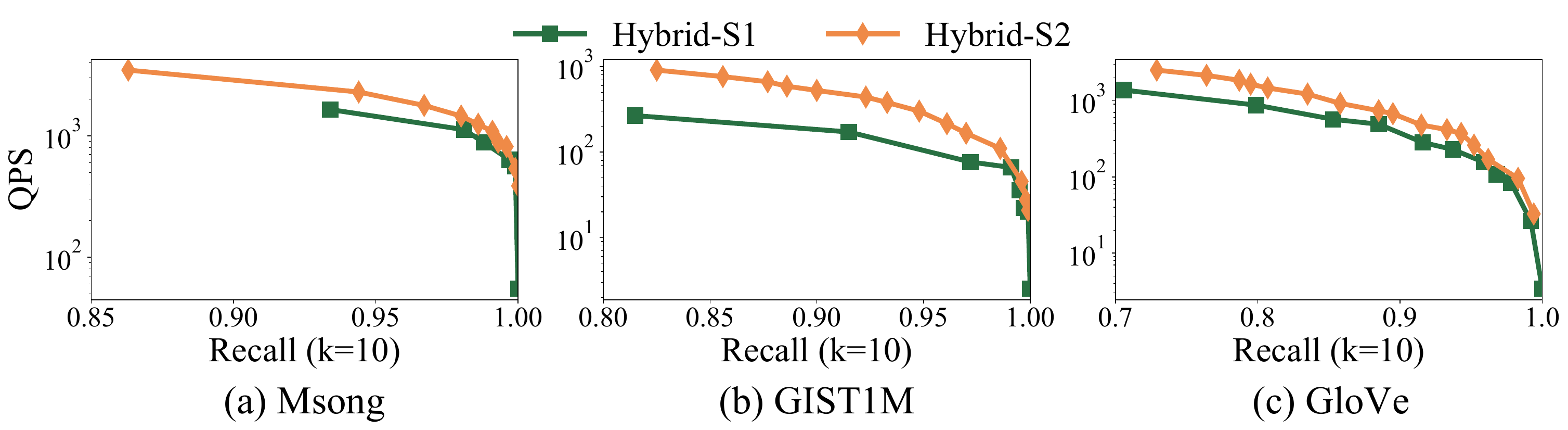}
  \caption{Impact of the runtime neighbor selection method.}
  \label{fig:alglobal}
\end{figure}

In this section, we compare two strategies for runtime neighbor selection in hybrid filtering, as described in \Cref{sec:hsig-search}. The first strategy, Hybrid-S1, computes the distance to neighbors in all $S^\prime$ segments (assuming there are $S^\prime$ segments intersecting with the query range) and then selects the top-$m$ neighbors by sorting them based on their distances. The second strategy, Hybrid-S2, selects the top-($\lceil m/{ S^\prime \rceil})$ neighbors from each segment. As shown in \Cref{fig:alglobal}, Hybrid-S2 outperforms Hybrid-S1 in query efficiency. This is because Hybrid-S2 selects the top-($\lceil m/{ S^\prime \rceil})$ neighbors from the pre-ordered neighbor lists without additional distance calculations, as neighbors of the object $v$ in each segment are already sorted by their distance to $v$ upon acquisition via \textit{ANNSearch}. In contrast, Hybrid-S1 requires calculating distances for all neighbors across $S^\prime$ segments, increasing query time. Therefore, we adopt Hybrid-S2 as the default runtime neighbor selection method.

\begin{figure}[!t]
  \centering
\includegraphics[width=\linewidth]{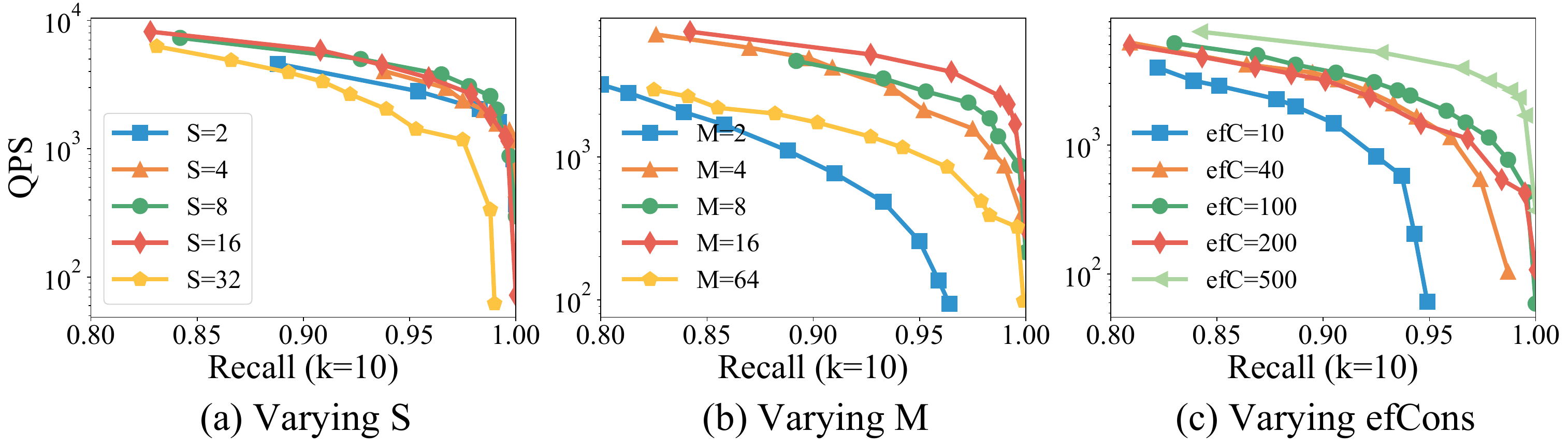}
\caption{Performance of index parameters on SIFT1M.}
  \label{fig:perform-with-bmef}
\end{figure}


\subsection{Parameter Study}
\begin{figure}[!t]
  \centering
\includegraphics[width=\linewidth]{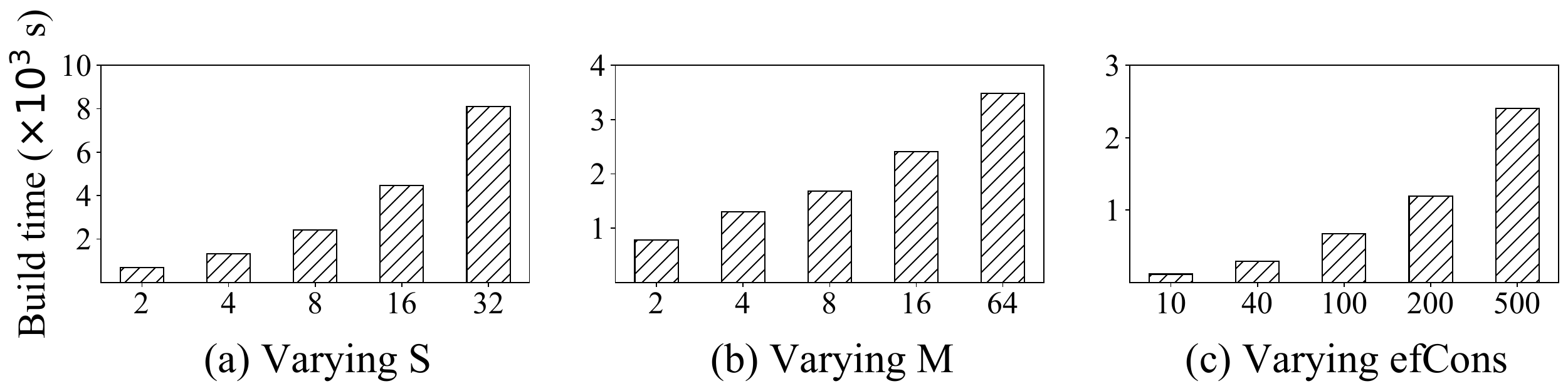}
  \caption{Index build time on different index parameters.}
  \label{fig:time-with-parameter}
\end{figure}

\noindent\textbf{Impact of Index Construction Parameters.}
We analyze the sensitivity of three parameters in \mbv construction: $S$, $M$, and $efCons$. 
\Cref{fig:perform-with-bmef,fig:time-with-parameter} show how parameters affect query performance and index build times, respectively. \Cref{fig:perform-with-bmef}a shows the impact of varying $S$ on query performance, with a performance increase from 2 to 8, followed by a decline as $S$ grows further due to increased neighbor visits. \Cref{fig:perform-with-bmef}b and \Cref{fig:perform-with-bmef}c illustrate the effects of varying $M$ and $efCons$ on query performance. Lower $M$ degrades graph quality, while higher $M$ increases the number of objects traversed during the search. Similarly, lower $efCons$ leads to insufficient candidates, whereas higher $efCons$ includes irrelevant candidates, both hindering query performance. Therefore, setting these parameters based on the dataset and workload is crucial for balancing query efficiency and accuracy.
As shown in \Cref{fig:time-with-parameter}, index build time rises with higher $S$, $M$, and $efCons$ due to increased distance computations. Based on a comprehensive evaluation of query performance and index construction time, we select $S=8$, $M=16$, and $efCons=500$ as default settings.

\begin{figure}[!t]
  \centering
\includegraphics[width=\linewidth]{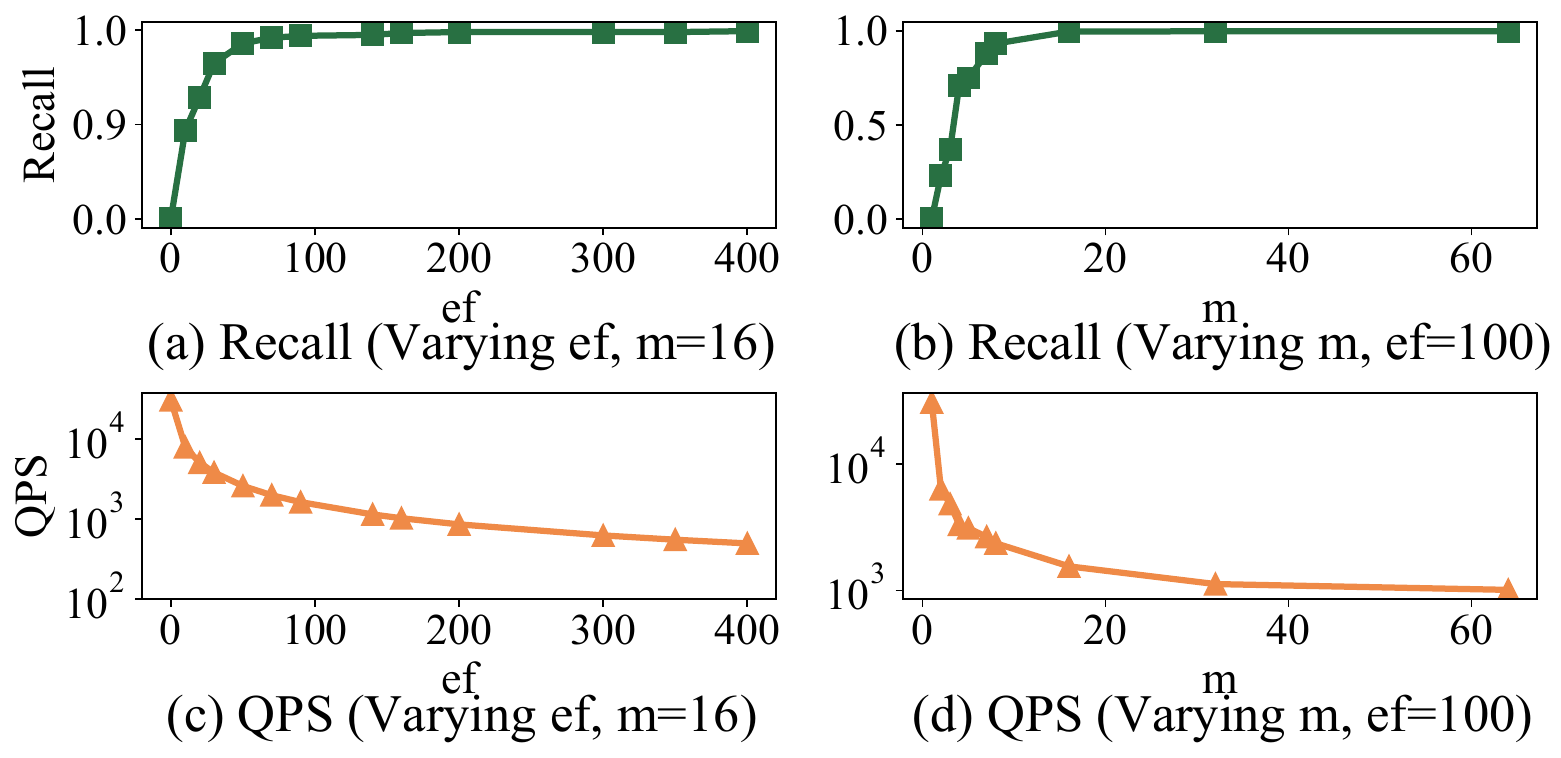}
\caption{Performance on different search parameters.}
  \label{fig:search parameter}
\end{figure}

\noindent\textbf{Impact of Search Parameters}. \Cref{fig:search parameter} shows the impact of varying the parameters $ef$ and $m$ on hybrid filtering in \mbv. Here, $ef$ is typically set to a value greater than $k$. \Cref{fig:search parameter}a and \Cref{fig:search parameter}c show that with $m=16$, as $ef$ increases, recall gradually improves while QPS decreases. This occurs because a larger $ef$ requires visiting more objects to gather sufficient candidates, enhancing accuracy but reducing efficiency. \Cref{fig:search parameter}b and \Cref{fig:search parameter}d show that with $ef=100$, increasing $m$ leads to a gradual improvement in recall but a decrease in query efficiency. This is because a larger $m$ results in visiting more neighbors of each object during the search, improving accuracy but at the cost of efficiency.

\begin{figure}[!t]
  \centering
\includegraphics[width=0.95\linewidth]{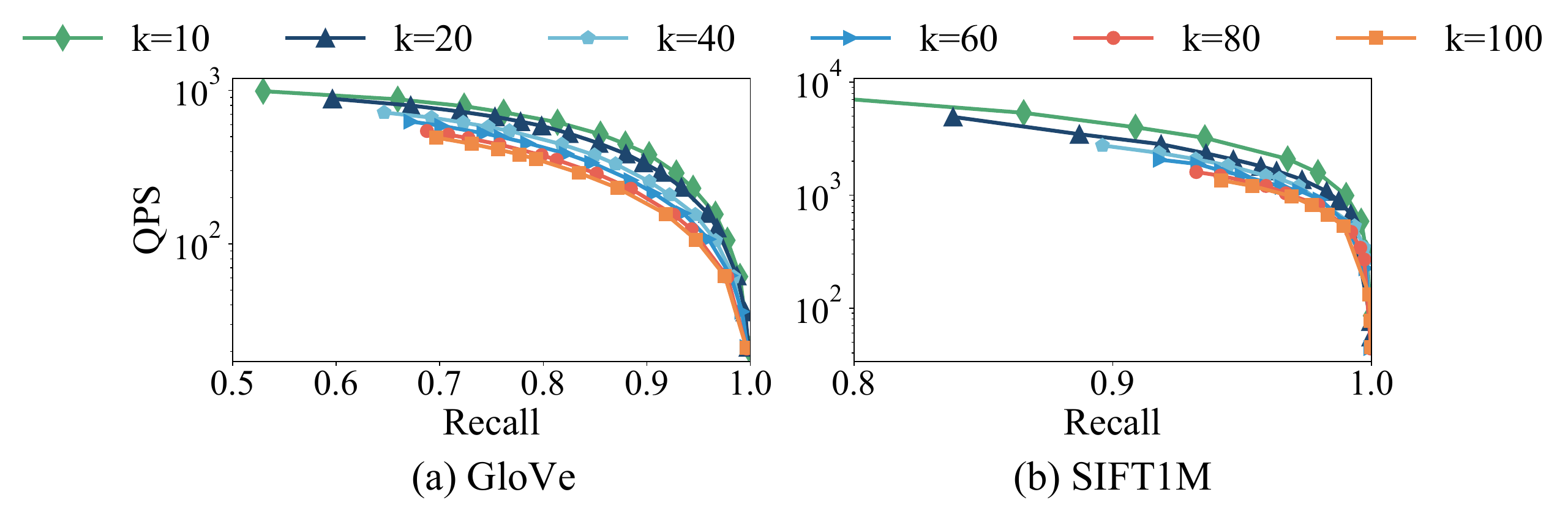}
\caption{Performance of different $k$ values.}
  \label{fig:diffkglove}
\end{figure}

\noindent\textbf{Impact of $k$ values}.
\Cref{fig:diffkglove} shows the impact of different $k$ values on query performance of \mbv. \mbv maintains strong efficiency and accuracy across various $k$ values. However, as $k$ increases from 10 to 100, performance gradually declines due to the increased number of candidates that need to be filtered during the search.

\subsection{Scalability} \label{sec:exp-scale}
\begin{figure}[!t]
  \centering
\includegraphics[width=\linewidth]{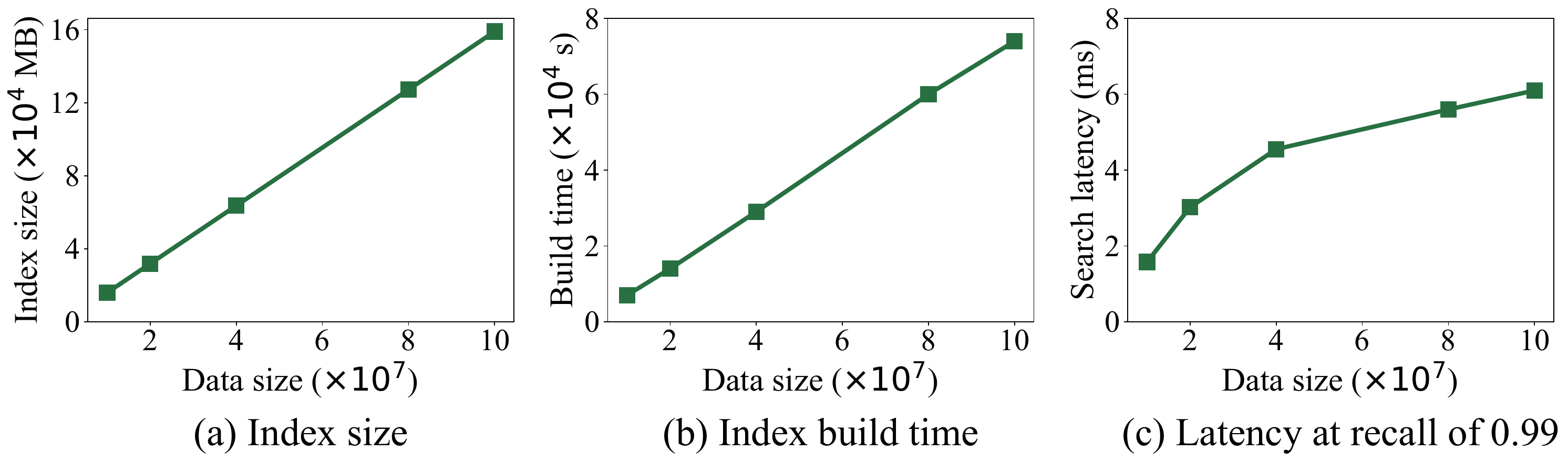}
\caption{Impact of varying data size on SIFT dataset.}
  \label{fig:scalable}
\end{figure}

We evaluate the scalability of \mbv using datasets ranging from 10 to 100 million objects. We fix the index parameters to $S=8$, $M=16$, and $efCons=500$, and maintain a query range width of 25\%. As shown in \Cref{fig:scalable}, both the index size and build time increase almost linearly with the dataset size. \Cref{fig:scalable}c plots the hybrid filtering latency versus data size, indicating a logarithmic search complexity. Notably, the recall consistently reaches 0.99 across all dataset sizes. These results demonstrate that \mbv achieves strong scalability in both index construction and query processing.

\subsection{Discussions}
\textbf{Range-aware strategy selection.} As mentioned in \Cref{sec:hsig-search}, our range-aware strategy selection method is based on historical data statistics, not query patterns. This approach may face challenges if the data distribution of the base dataset changes significantly over time. To address this issue, we propose an adaptive method that can detect changes in data distribution. If the change exceeds the user-defined threshold, the method resamples objects from the updated base dataset to recalibrate $\tau_A$ and $\tau_B$, thereby adjusting the range-aware search strategy selection.

\noindent\textbf{\rann with Multiple Attributes.} Existing PG-based indexes struggle to support \rann queries with multiple attributes, as incorporating multiple attributes into a graph is challenging. However, with two enhancements, \mbv can be extended to handle such queries. While we use the case with two attributes in this discussion, the method can be adapted for more attributes. (1) Multiple single-attribute indexes: Build a separate \mbv for each attribute. For a conjunctive query with a query vector $q$, retrieve objects that satisfy $r_1(A_1)$ AND $r_2(A_2)$, where $r_1(A_1)$ and $r_2(A_2)$ are the attribute ranges. We aim to return the top-$k$ ANN of $q$ among objects satisfying both ranges. Specifically, we modify Line 15 of Algorithm 7 to "if $v$ satisfies $r_1(A_1)$ AND $r_2(A_2)$, then push $v$ to $ann$". For disjunctive queries ($r_1(A_1)$ OR $r_2(A_2)$), we use separate indexes for $A_1$ and $A_2$ and merge the results. (2) Single index for multiple attributes: Create a composite attribute $(A_1, A_2)$ and apply the z-order method \cite{orenstein1984class} to map them into a one-dimensional attribute, enabling the construction of a single \mbv to handle \rann queries with multiple attributes. We plan to explore a dedicated algorithm for \rann queries with multiple attributes in future work.

\section{RELATED WORK}\label{sec:related work}
\subsection{ANNS}
The primary approaches for A$k$NNS can be categorized tree-based methods \cite{arora2018hd,muja2014scalable}, hash-based methods \cite{huang2015query,zheng2020pm,tian2023db}, quantization-based methods \cite{andre2016cache,ge2013optimized,liu2017pqbf,matsui2015pqtable}, and PG-based methods \cite{nsg,malkov2018efficient,fu2021high,wang2021comprehensive,ren2020hm,song2024efficient,zhang2023learning}. Tree-structured indexes like the KD-tree \cite{kd-tree}, R-tree \cite{guttman1984r}, VP-tree \cite{vp-tree}, and KMeans-tree \cite{yianilos1993data} suffer from the "curse of dimensionality" \cite{indyk1998approximate}, making them ineffective in high-dimensional spaces. Hash-based methods utilize hash functions to map vectors into hash buckets. However, as the binary hash code length increases, the number of buckets grows exponentially, leading to many empty buckets, which reduces the search accuracy. Quantization-based methods reduce storage and computational costs but involve lossy compression, which produces a "ceiling" phenomenon on the search accuracy \cite{li2019approximate}. PG-based methods show significant performance advantages and have attracted substantial attention. However, while effective for vector retrieval, these methods fail to handle attribute filtering effectively, limiting their applicability in scenarios requiring integrated vector retrieval and attribute filtering.

\subsection{Filtered ANNS}
Most hybrid A$k$NNS queries separate the process into vector retrieval and attribute filtering, which are combined to produce final results. MA-NSW \cite{xu2020multiattribute} explores ANNS with attribute constraints by constructing indexes for each attribute combination. \vear \cite{li2018design} and \ngt \cite{NGTonline} apply post-filtering, which first retrieves candidates through vector search and then filters candidates based on the attributes. This strategy is extendable to some vector libraries, such as Faiss \cite{johnson2019billion} and SPTAG \cite{sptag}. However, they perform worse when the selectivity of the query range is low, limiting the query efficiency and accuracy.
\adb \cite{wei2020analyticdb} uses a B-tree for attributes and a PQ index for vectors, optimizing query plans with a cost model. \mil \cite{wang2021milvus} partitions datasets by attributes and adopts the query strategies of \adb. However, they focus on query optimization and partitioning techniques without enhancing the index structure.
Filtered-DiskANN \cite{gollapudi2023filtered} develops a graph supporting attribute matching and vector similarity searches. However, it focuses on attribute matching, leaving a gap for A$k$NN with range constraints. \nhq \cite{wang2024efficient} and HQANN \cite{wu2022hqann} introduce a fused distance metric that combines attributes and vectors, enabling simultaneous attribute filtering and vector retrieval within a single graph index. However, they lack a solid theoretical foundation due to the irrelevance of attributes and vectors. ARKGraph \cite{zuo2023arkgraph} builds PGs for all possible attribute range combinations and compresses the indexes. However, it requires decompression during querying, reducing query efficiency. SeRF \cite{zuo2024serf} addresses range-filtering A$k$NNS by designing a segment graph that compresses multiple indexes for half-bounded range queries and extends it to support general range queries. However, SeRF does not support online updates of new data.

\section{CONCLUSION}\label{sec:conclusion}
This paper addresses \rann queries over high-dimensional vectors associated with attribute values. Existing methods, including pre-, post-, and hybrid filtering strategies, which apply attribute filtering before, after, or during the ANNS process, suffer performance degradation when query ranges shift. We propose a novel framework called UNIFY, which constructs a unified PG-based index that seamlessly supports all three strategies. Within UNIFY, we introduce SIG, enabling efficient \rann by reconstructing and searching a PG from relevant segments. Additionally, we present \mbv, a variant of SIG that incorporates a hierarchical structure inspired by HNSW, achieving logarithmic time complexity for \rann. Experimental results demonstrate that UNIFY outperforms state-of-the-art methods across varying query ranges.

\begin{acks}
 Bin Yao was supported by the Oceanic Interdisciplinary Program of Shanghai Jiao Tong University (SL2023ZD102) and Alibaba Group through Alibaba Innovative Research (AIR) Program. Zhongpu Chen was supported by Sichuan Science and Technology Program (2024NSFSC1460).
\end{acks}


\bibliographystyle{ACM-Reference-Format}
\balance
\bibliography{sample}


\begin{thebibliography}{52}


\ifx \showCODEN    \undefined \def \showCODEN     #1{\unskip}     \fi
\ifx \showDOI      \undefined \def \showDOI       #1{#1}\fi
\ifx \showISBNx    \undefined \def \showISBNx     #1{\unskip}     \fi
\ifx \showISBNxiii \undefined \def \showISBNxiii  #1{\unskip}     \fi
\ifx \showISSN     \undefined \def \showISSN      #1{\unskip}     \fi
\ifx \showLCCN     \undefined \def \showLCCN      #1{\unskip}     \fi
\ifx \shownote     \undefined \def \shownote      #1{#1}          \fi
\ifx \showarticletitle \undefined \def \showarticletitle #1{#1}   \fi
\ifx \showURL      \undefined \def \showURL       {\relax}        \fi
\providecommand\bibfield[2]{#2}
\providecommand\bibinfo[2]{#2}
\providecommand\natexlab[1]{#1}
\providecommand\showeprint[2][]{arXiv:#2}

\bibitem[\protect\citeauthoryear{Andr{\'e}, Kermarrec, and Le~Scouarnec}{Andr{\'e} et~al\mbox{.}}{2016}]%
        {andre2016cache}
\bibfield{author}{\bibinfo{person}{Fabien Andr{\'e}}, \bibinfo{person}{Anne-Marie Kermarrec}, {and} \bibinfo{person}{Nicolas Le~Scouarnec}.} \bibinfo{year}{2016}\natexlab{}.
\newblock \showarticletitle{Cache locality is not enough: High-performance nearest neighbor search with product quantization fast scan}. In \bibinfo{booktitle}{\emph{VLDB}}. \bibinfo{pages}{288--299}.
\newblock


\bibitem[\protect\citeauthoryear{Arora, Sinha, Kumar, and Bhattacharya}{Arora et~al\mbox{.}}{2018}]%
        {arora2018hd}
\bibfield{author}{\bibinfo{person}{Akhil Arora}, \bibinfo{person}{Sakshi Sinha}, \bibinfo{person}{Piyush Kumar}, {and} \bibinfo{person}{Arnab Bhattacharya}.} \bibinfo{year}{2018}\natexlab{}.
\newblock \showarticletitle{Hd-index: Pushing the scalability-accuracy boundary for approximate knn search in high-dimensional spaces}. In \bibinfo{booktitle}{\emph{VLDB}}. \bibinfo{pages}{906--919}.
\newblock


\bibitem[\protect\citeauthoryear{Aum{\"u}ller, Bernhardsson, and Faithfull}{Aum{\"u}ller et~al\mbox{.}}{2020}]%
        {aumuller2020ann}
\bibfield{author}{\bibinfo{person}{Martin Aum{\"u}ller}, \bibinfo{person}{Erik Bernhardsson}, {and} \bibinfo{person}{Alexander Faithfull}.} \bibinfo{year}{2020}\natexlab{}.
\newblock \showarticletitle{ANN-Benchmarks: A benchmarking tool for approximate nearest neighbor algorithms}.
\newblock \bibinfo{journal}{\emph{Information Systems}}  \bibinfo{volume}{87} (\bibinfo{year}{2020}), \bibinfo{pages}{101374}.
\newblock


\bibitem[\protect\citeauthoryear{Bentley}{Bentley}{1975}]%
        {kd-tree}
\bibfield{author}{\bibinfo{person}{Jon~Louis Bentley}.} \bibinfo{year}{1975}\natexlab{}.
\newblock \showarticletitle{Multidimensional binary search trees used for associative searching}.
\newblock \bibinfo{journal}{\emph{Commun. ACM}}  \bibinfo{volume}{18} (\bibinfo{year}{1975}), \bibinfo{pages}{509--517}.
\newblock


\bibitem[\protect\citeauthoryear{Chen, Wang, Li, Ren, Li, Zhu, Li, Liu, Zhang, and Wang}{Chen et~al\mbox{.}}{2018}]%
        {sptag}
\bibfield{author}{\bibinfo{person}{Qi Chen}, \bibinfo{person}{Haidong Wang}, \bibinfo{person}{Mingqin Li}, \bibinfo{person}{Gang Ren}, \bibinfo{person}{Scarlett Li}, \bibinfo{person}{Jeffery Zhu}, \bibinfo{person}{Jason Li}, \bibinfo{person}{Chuanjie Liu}, \bibinfo{person}{Lintao Zhang}, {and} \bibinfo{person}{Jingdong Wang}.} \bibinfo{year}{2018}\natexlab{}.
\newblock \bibinfo{booktitle}{\emph{SPTAG: A library for fast approximate nearest neighbor search}}.
\newblock
\urldef\tempurl%
\url{https://github.com/Microsoft/SPTAG.}
\showURL{%
\tempurl}


\bibitem[\protect\citeauthoryear{Fu, Wang, and Cai}{Fu et~al\mbox{.}}{2021}]%
        {fu2021high}
\bibfield{author}{\bibinfo{person}{Cong Fu}, \bibinfo{person}{Changxu Wang}, {and} \bibinfo{person}{Deng Cai}.} \bibinfo{year}{2021}\natexlab{}.
\newblock \showarticletitle{High dimensional similarity search with satellite system graph: Efficiency, scalability, and unindexed query compatibility}.
\newblock \bibinfo{journal}{\emph{TPAMI}} \bibinfo{volume}{44}, \bibinfo{number}{8} (\bibinfo{year}{2021}), \bibinfo{pages}{4139--4150}.
\newblock


\bibitem[\protect\citeauthoryear{Fu, Xiang, Wang, and Cai}{Fu et~al\mbox{.}}{2019}]%
        {nsg}
\bibfield{author}{\bibinfo{person}{Cong Fu}, \bibinfo{person}{Chao Xiang}, \bibinfo{person}{Changxu Wang}, {and} \bibinfo{person}{Deng Cai}.} \bibinfo{year}{2019}\natexlab{}.
\newblock \showarticletitle{Fast approximate nearest neighbor search with the navigating spreading-out graph}. In \bibinfo{booktitle}{\emph{VLDB}}. \bibinfo{pages}{461--474}.
\newblock


\bibitem[\protect\citeauthoryear{Fukunaga and Narendra}{Fukunaga and Narendra}{1975}]%
        {vp-tree}
\bibfield{author}{\bibinfo{person}{Keinosuke Fukunaga} {and} \bibinfo{person}{Patrenahalli~M. Narendra}.} \bibinfo{year}{1975}\natexlab{}.
\newblock \showarticletitle{A branch and bound algorithm for computing k-nearest neighbors}.
\newblock \bibinfo{journal}{\emph{IEEE transactions on computers}} \bibinfo{volume}{100}, \bibinfo{number}{7} (\bibinfo{year}{1975}), \bibinfo{pages}{750--753}.
\newblock


\bibitem[\protect\citeauthoryear{Ge, He, Ke, and Sun}{Ge et~al\mbox{.}}{2013}]%
        {ge2013optimized}
\bibfield{author}{\bibinfo{person}{Tiezheng Ge}, \bibinfo{person}{Kaiming He}, \bibinfo{person}{Qifa Ke}, {and} \bibinfo{person}{Jian Sun}.} \bibinfo{year}{2013}\natexlab{}.
\newblock \showarticletitle{Optimized product quantization for approximate nearest neighbor search}. In \bibinfo{booktitle}{\emph{CVPR}}. \bibinfo{pages}{2946--2953}.
\newblock


\bibitem[\protect\citeauthoryear{Gollapudi, Karia, Sivashankar, Krishnaswamy, Begwani, Raz, Lin, Zhang, Mahapatro, Srinivasan, et~al\mbox{.}}{Gollapudi et~al\mbox{.}}{2023}]%
        {gollapudi2023filtered}
\bibfield{author}{\bibinfo{person}{Siddharth Gollapudi}, \bibinfo{person}{Neel Karia}, \bibinfo{person}{Varun Sivashankar}, \bibinfo{person}{Ravishankar Krishnaswamy}, \bibinfo{person}{Nikit Begwani}, \bibinfo{person}{Swapnil Raz}, \bibinfo{person}{Yiyong Lin}, \bibinfo{person}{Yin Zhang}, \bibinfo{person}{Neelam Mahapatro}, \bibinfo{person}{Premkumar Srinivasan}, {et~al\mbox{.}}} \bibinfo{year}{2023}\natexlab{}.
\newblock \showarticletitle{Filtered-diskann: Graph algorithms for approximate nearest neighbor search with filters}. In \bibinfo{booktitle}{\emph{Proceedings of the ACM Web Conference 2023}}. \bibinfo{pages}{3406--3416}.
\newblock


\bibitem[\protect\citeauthoryear{Guttman}{Guttman}{1984}]%
        {guttman1984r}
\bibfield{author}{\bibinfo{person}{Antonin Guttman}.} \bibinfo{year}{1984}\natexlab{}.
\newblock \showarticletitle{R-trees: A dynamic index structure for spatial searching}. In \bibinfo{booktitle}{\emph{SIGMOD}}. \bibinfo{pages}{47--57}.
\newblock


\bibitem[\protect\citeauthoryear{Harmouch and Naumann}{Harmouch and Naumann}{2017}]%
        {harmouch2017cardinality}
\bibfield{author}{\bibinfo{person}{Hazar Harmouch} {and} \bibinfo{person}{Felix Naumann}.} \bibinfo{year}{2017}\natexlab{}.
\newblock \showarticletitle{Cardinality estimation: An experimental survey}. In \bibinfo{booktitle}{\emph{VLDB}}. \bibinfo{pages}{499--512}.
\newblock


\bibitem[\protect\citeauthoryear{Huang, Feng, Zhang, Fang, and Ng}{Huang et~al\mbox{.}}{2015}]%
        {huang2015query}
\bibfield{author}{\bibinfo{person}{Qiang Huang}, \bibinfo{person}{Jianlin Feng}, \bibinfo{person}{Yikai Zhang}, \bibinfo{person}{Qiong Fang}, {and} \bibinfo{person}{Wilfred Ng}.} \bibinfo{year}{2015}\natexlab{}.
\newblock \showarticletitle{Query-aware locality-sensitive hashing for approximate nearest neighbor search}. In \bibinfo{booktitle}{\emph{VLDB}}. \bibinfo{pages}{1--12}.
\newblock


\bibitem[\protect\citeauthoryear{Indyk and Motwani}{Indyk and Motwani}{1998}]%
        {indyk1998approximate}
\bibfield{author}{\bibinfo{person}{Piotr Indyk} {and} \bibinfo{person}{Rajeev Motwani}.} \bibinfo{year}{1998}\natexlab{}.
\newblock \showarticletitle{Approximate nearest neighbors: towards removing the curse of dimensionality}. In \bibinfo{booktitle}{\emph{Proceedings of the thirtieth annual ACM symposium on Theory of computing}}. \bibinfo{pages}{604--613}.
\newblock


\bibitem[\protect\citeauthoryear{Jayaram~Subramanya, Devvrit, Simhadri, Krishnawamy, and Kadekodi}{Jayaram~Subramanya et~al\mbox{.}}{2019}]%
        {jayaram2019diskann}
\bibfield{author}{\bibinfo{person}{Suhas Jayaram~Subramanya}, \bibinfo{person}{Fnu Devvrit}, \bibinfo{person}{Harsha~Vardhan Simhadri}, \bibinfo{person}{Ravishankar Krishnawamy}, {and} \bibinfo{person}{Rohan Kadekodi}.} \bibinfo{year}{2019}\natexlab{}.
\newblock \showarticletitle{Diskann: Fast accurate billion-point nearest neighbor search on a single node}. In \bibinfo{booktitle}{\emph{NeurIPS}}, Vol.~\bibinfo{volume}{32}.
\newblock


\bibitem[\protect\citeauthoryear{Jegou, Douze, and Schmid}{Jegou et~al\mbox{.}}{2010}]%
        {jegou2010product}
\bibfield{author}{\bibinfo{person}{Herve Jegou}, \bibinfo{person}{Matthijs Douze}, {and} \bibinfo{person}{Cordelia Schmid}.} \bibinfo{year}{2010}\natexlab{}.
\newblock \showarticletitle{Product quantization for nearest neighbor search}.
\newblock \bibinfo{journal}{\emph{IEEE transactions on pattern analysis and machine intelligence}} \bibinfo{volume}{33}, \bibinfo{number}{1} (\bibinfo{year}{2010}), \bibinfo{pages}{117--128}.
\newblock


\bibitem[\protect\citeauthoryear{Jingdong}{Jingdong}{2020}]%
        {vearchcode}
\bibfield{author}{\bibinfo{person}{Jingdong}.} \bibinfo{year}{2020}\natexlab{}.
\newblock \bibinfo{booktitle}{\emph{A distributed system for embedding-based retrieval}}.
\newblock
\urldef\tempurl%
\url{https://github.com/vearch/vearch.}
\showURL{%
\tempurl}


\bibitem[\protect\citeauthoryear{Johnson, Douze, and J{\'e}gou}{Johnson et~al\mbox{.}}{2019}]%
        {johnson2019billion}
\bibfield{author}{\bibinfo{person}{Jeff Johnson}, \bibinfo{person}{Matthijs Douze}, {and} \bibinfo{person}{Herv{\'e} J{\'e}gou}.} \bibinfo{year}{2019}\natexlab{}.
\newblock \showarticletitle{Billion-scale similarity search with GPUs}.
\newblock \bibinfo{journal}{\emph{IEEE Transactions on Big Data}} \bibinfo{volume}{7}, \bibinfo{number}{3} (\bibinfo{year}{2019}), \bibinfo{pages}{535--547}.
\newblock


\bibitem[\protect\citeauthoryear{Lewis, Perez, Piktus, Petroni, Karpukhin, Goyal, K{\"u}ttler, Lewis, Yih, Rockt{\"a}schel, et~al\mbox{.}}{Lewis et~al\mbox{.}}{2020}]%
        {lewis2020retrieval}
\bibfield{author}{\bibinfo{person}{Patrick Lewis}, \bibinfo{person}{Ethan Perez}, \bibinfo{person}{Aleksandra Piktus}, \bibinfo{person}{Fabio Petroni}, \bibinfo{person}{Vladimir Karpukhin}, \bibinfo{person}{Naman Goyal}, \bibinfo{person}{Heinrich K{\"u}ttler}, \bibinfo{person}{Mike Lewis}, \bibinfo{person}{Wen-tau Yih}, \bibinfo{person}{Tim Rockt{\"a}schel}, {et~al\mbox{.}}} \bibinfo{year}{2020}\natexlab{}.
\newblock \showarticletitle{Retrieval-augmented generation for knowledge-intensive nlp tasks}.
\newblock \bibinfo{journal}{\emph{Advances in Neural Information Processing Systems}}  \bibinfo{volume}{33} (\bibinfo{year}{2020}), \bibinfo{pages}{9459--9474}.
\newblock


\bibitem[\protect\citeauthoryear{Li, Liu, Gui, Chen, Ni, Wang, and Chen}{Li et~al\mbox{.}}{2018}]%
        {li2018design}
\bibfield{author}{\bibinfo{person}{Jie Li}, \bibinfo{person}{Haifeng Liu}, \bibinfo{person}{Chuanghua Gui}, \bibinfo{person}{Jianyu Chen}, \bibinfo{person}{Zhenyuan Ni}, \bibinfo{person}{Ning Wang}, {and} \bibinfo{person}{Yuan Chen}.} \bibinfo{year}{2018}\natexlab{}.
\newblock \showarticletitle{The design and implementation of a real time visual search system on JD E-commerce platform}. In \bibinfo{booktitle}{\emph{Proceedings of the 19th International Middleware Conference Industry}}. \bibinfo{pages}{9--16}.
\newblock


\bibitem[\protect\citeauthoryear{Li, Zhang, Sun, Wang, Li, Zhang, and Lin}{Li et~al\mbox{.}}{2019}]%
        {li2019approximate}
\bibfield{author}{\bibinfo{person}{Wen Li}, \bibinfo{person}{Ying Zhang}, \bibinfo{person}{Yifang Sun}, \bibinfo{person}{Wei Wang}, \bibinfo{person}{Mingjie Li}, \bibinfo{person}{Wenjie Zhang}, {and} \bibinfo{person}{Xuemin Lin}.} \bibinfo{year}{2019}\natexlab{}.
\newblock \showarticletitle{Approximate nearest neighbor search on high dimensional data—experiments, analyses, and improvement}.
\newblock \bibinfo{journal}{\emph{TKDE}} \bibinfo{volume}{32}, \bibinfo{number}{8} (\bibinfo{year}{2019}), \bibinfo{pages}{1475--1488}.
\newblock


\bibitem[\protect\citeauthoryear{Liu, Cheng, and Cui}{Liu et~al\mbox{.}}{2017}]%
        {liu2017pqbf}
\bibfield{author}{\bibinfo{person}{Yingfan Liu}, \bibinfo{person}{Hong Cheng}, {and} \bibinfo{person}{Jiangtao Cui}.} \bibinfo{year}{2017}\natexlab{}.
\newblock \showarticletitle{PQBF: i/o-efficient approximate nearest neighbor search by product quantization}. In \bibinfo{booktitle}{\emph{CIKM}}. \bibinfo{pages}{667--676}.
\newblock


\bibitem[\protect\citeauthoryear{Malkov and Yashunin}{Malkov and Yashunin}{2018}]%
        {malkov2018efficient}
\bibfield{author}{\bibinfo{person}{Yu~A Malkov} {and} \bibinfo{person}{Dmitry~A Yashunin}.} \bibinfo{year}{2018}\natexlab{}.
\newblock \showarticletitle{Efficient and robust approximate nearest neighbor search using hierarchical navigable small world graphs}.
\newblock \bibinfo{journal}{\emph{TPAMI}} \bibinfo{volume}{42}, \bibinfo{number}{4} (\bibinfo{year}{2018}), \bibinfo{pages}{824--836}.
\newblock


\bibitem[\protect\citeauthoryear{Matsui, Yamasaki, and Aizawa}{Matsui et~al\mbox{.}}{2015}]%
        {matsui2015pqtable}
\bibfield{author}{\bibinfo{person}{Yusuke Matsui}, \bibinfo{person}{Toshihiko Yamasaki}, {and} \bibinfo{person}{Kiyoharu Aizawa}.} \bibinfo{year}{2015}\natexlab{}.
\newblock \showarticletitle{Pqtable: Fast exact asymmetric distance neighbor search for product quantization using hash tables}. In \bibinfo{booktitle}{\emph{ICCV}}. \bibinfo{pages}{1940--1948}.
\newblock


\bibitem[\protect\citeauthoryear{Mo, Peng, Xu, Shi, and Zhu}{Mo et~al\mbox{.}}{2022}]%
        {mo2022simple}
\bibfield{author}{\bibinfo{person}{Yujie Mo}, \bibinfo{person}{Liang Peng}, \bibinfo{person}{Jie Xu}, \bibinfo{person}{Xiaoshuang Shi}, {and} \bibinfo{person}{Xiaofeng Zhu}.} \bibinfo{year}{2022}\natexlab{}.
\newblock \showarticletitle{Simple unsupervised graph representation learning}. In \bibinfo{booktitle}{\emph{AAAI}}. \bibinfo{pages}{7797--7805}.
\newblock


\bibitem[\protect\citeauthoryear{Muja and Lowe}{Muja and Lowe}{2014}]%
        {muja2014scalable}
\bibfield{author}{\bibinfo{person}{Marius Muja} {and} \bibinfo{person}{David~G Lowe}.} \bibinfo{year}{2014}\natexlab{}.
\newblock \showarticletitle{Scalable nearest neighbor algorithms for high dimensional data}.
\newblock \bibinfo{journal}{\emph{TPAMI}} \bibinfo{volume}{36}, \bibinfo{number}{11} (\bibinfo{year}{2014}), \bibinfo{pages}{2227--2240}.
\newblock


\bibitem[\protect\citeauthoryear{Oommen and Rueda}{Oommen and Rueda}{2002}]%
        {oommen2002efficiency}
\bibfield{author}{\bibinfo{person}{B~John Oommen} {and} \bibinfo{person}{Luis~G Rueda}.} \bibinfo{year}{2002}\natexlab{}.
\newblock \showarticletitle{The efficiency of histogram-like techniques for database query optimization}.
\newblock \bibinfo{journal}{\emph{Comput. J.}} \bibinfo{volume}{45}, \bibinfo{number}{5} (\bibinfo{year}{2002}), \bibinfo{pages}{494--510}.
\newblock


\bibitem[\protect\citeauthoryear{Orenstein and Merrett}{Orenstein and Merrett}{1984}]%
        {orenstein1984class}
\bibfield{author}{\bibinfo{person}{Jack~A Orenstein} {and} \bibinfo{person}{Tim~H Merrett}.} \bibinfo{year}{1984}\natexlab{}.
\newblock \showarticletitle{A class of data structures for associative searching}. In \bibinfo{booktitle}{\emph{Proceedings of the 3rd ACM SIGACT-SIGMOD Symposium on Principles of Database Systems}}. \bibinfo{pages}{181--190}.
\newblock


\bibitem[\protect\citeauthoryear{Paredes, Ch{\'a}vez, Figueroa, and Navarro}{Paredes et~al\mbox{.}}{2006}]%
        {paredes2006practical}
\bibfield{author}{\bibinfo{person}{Rodrigo Paredes}, \bibinfo{person}{Edgar Ch{\'a}vez}, \bibinfo{person}{Karina Figueroa}, {and} \bibinfo{person}{Gonzalo Navarro}.} \bibinfo{year}{2006}\natexlab{}.
\newblock \showarticletitle{Practical construction of k-nearest neighbor graphs in metric spaces}. In \bibinfo{booktitle}{\emph{International Workshop on Experimental and Efficient Algorithms}}. \bibinfo{pages}{85--97}.
\newblock


\bibitem[\protect\citeauthoryear{Piatetsky-Shapiro and Connell}{Piatetsky-Shapiro and Connell}{1984}]%
        {piatetsky1984accurate}
\bibfield{author}{\bibinfo{person}{Gregory Piatetsky-Shapiro} {and} \bibinfo{person}{Charles Connell}.} \bibinfo{year}{1984}\natexlab{}.
\newblock \showarticletitle{Accurate estimation of the number of tuples satisfying a condition}. In \bibinfo{booktitle}{\emph{SIGMOD}}. \bibinfo{pages}{256--276}.
\newblock


\bibitem[\protect\citeauthoryear{Pinecone}{Pinecone}{2021}]%
        {pinecone}
\bibfield{author}{\bibinfo{person}{Pinecone}.} \bibinfo{year}{2021}\natexlab{}.
\newblock \bibinfo{booktitle}{\emph{Pinecone.io}}.
\newblock
\urldef\tempurl%
\url{https://www.pinecone.io/.}
\showURL{%
\tempurl}


\bibitem[\protect\citeauthoryear{Pugh}{Pugh}{1990}]%
        {pugh1990skip}
\bibfield{author}{\bibinfo{person}{William Pugh}.} \bibinfo{year}{1990}\natexlab{}.
\newblock \showarticletitle{Skip lists: a probabilistic alternative to balanced trees}.
\newblock \bibinfo{journal}{\emph{Commun. ACM}} \bibinfo{volume}{33}, \bibinfo{number}{6} (\bibinfo{year}{1990}), \bibinfo{pages}{668--676}.
\newblock


\bibitem[\protect\citeauthoryear{Qin, Wang, Xiao, and Zhang}{Qin et~al\mbox{.}}{2020}]%
        {qin2020similarity}
\bibfield{author}{\bibinfo{person}{Jianbin Qin}, \bibinfo{person}{Wei Wang}, \bibinfo{person}{Chuan Xiao}, {and} \bibinfo{person}{Ying Zhang}.} \bibinfo{year}{2020}\natexlab{}.
\newblock \showarticletitle{Similarity query processing for high-dimensional data}. In \bibinfo{booktitle}{\emph{VLDB}}. \bibinfo{pages}{3437--3440}.
\newblock


\bibitem[\protect\citeauthoryear{Ren, Zhang, and Li}{Ren et~al\mbox{.}}{2020}]%
        {ren2020hm}
\bibfield{author}{\bibinfo{person}{Jie Ren}, \bibinfo{person}{Minjia Zhang}, {and} \bibinfo{person}{Dong Li}.} \bibinfo{year}{2020}\natexlab{}.
\newblock \showarticletitle{Hm-ann: Efficient billion-point nearest neighbor search on heterogeneous memory}. In \bibinfo{booktitle}{\emph{NeurIPS}}. \bibinfo{pages}{10672--10684}.
\newblock


\bibitem[\protect\citeauthoryear{Song, Wang, Yao, Chen, Xie, and Li}{Song et~al\mbox{.}}{2024}]%
        {song2024efficient}
\bibfield{author}{\bibinfo{person}{Yitong Song}, \bibinfo{person}{Kai Wang}, \bibinfo{person}{Bin Yao}, \bibinfo{person}{Zhida Chen}, \bibinfo{person}{Jiong Xie}, {and} \bibinfo{person}{Feifei Li}.} \bibinfo{year}{2024}\natexlab{}.
\newblock \showarticletitle{Efficient Reverse $ k $ Approximate Nearest Neighbor Search Over High-Dimensional Vectors}. In \bibinfo{booktitle}{\emph{ICDE}}. \bibinfo{pages}{4262--4274}.
\newblock


\bibitem[\protect\citeauthoryear{Suchal and N{\'a}vrat}{Suchal and N{\'a}vrat}{2010}]%
        {suchal2010full}
\bibfield{author}{\bibinfo{person}{J{\'a}n Suchal} {and} \bibinfo{person}{Pavol N{\'a}vrat}.} \bibinfo{year}{2010}\natexlab{}.
\newblock \showarticletitle{Full text search engine as scalable k-nearest neighbor recommendation system}. In \bibinfo{booktitle}{\emph{IFIP International Conference on Artificial Intelligence}}. \bibinfo{pages}{165--173}.
\newblock


\bibitem[\protect\citeauthoryear{Tian, Zhao, and Zhou}{Tian et~al\mbox{.}}{2023}]%
        {tian2023db}
\bibfield{author}{\bibinfo{person}{Yao Tian}, \bibinfo{person}{Xi Zhao}, {and} \bibinfo{person}{Xiaofang Zhou}.} \bibinfo{year}{2023}\natexlab{}.
\newblock \showarticletitle{DB-LSH 2.0: Locality-Sensitive Hashing With Query-Based Dynamic Bucketing}.
\newblock \bibinfo{journal}{\emph{TKDE}} \bibinfo{volume}{36}, \bibinfo{number}{3} (\bibinfo{year}{2023}), \bibinfo{pages}{1000--1015}.
\newblock


\bibitem[\protect\citeauthoryear{Valkanas, Lappas, and Gunopulos}{Valkanas et~al\mbox{.}}{2017}]%
        {valkanas2017mining}
\bibfield{author}{\bibinfo{person}{George Valkanas}, \bibinfo{person}{Theodoros Lappas}, {and} \bibinfo{person}{Dimitrios Gunopulos}.} \bibinfo{year}{2017}\natexlab{}.
\newblock \showarticletitle{Mining competitors from large unstructured datasets}.
\newblock \bibinfo{journal}{\emph{TKDE}} \bibinfo{volume}{29}, \bibinfo{number}{9} (\bibinfo{year}{2017}), \bibinfo{pages}{1971--1984}.
\newblock


\bibitem[\protect\citeauthoryear{Wang, Yi, Guo, Jin, Xu, Li, Wang, Guo, Li, Xu, et~al\mbox{.}}{Wang et~al\mbox{.}}{2021b}]%
        {wang2021milvus}
\bibfield{author}{\bibinfo{person}{Jianguo Wang}, \bibinfo{person}{Xiaomeng Yi}, \bibinfo{person}{Rentong Guo}, \bibinfo{person}{Hai Jin}, \bibinfo{person}{Peng Xu}, \bibinfo{person}{Shengjun Li}, \bibinfo{person}{Xiangyu Wang}, \bibinfo{person}{Xiangzhou Guo}, \bibinfo{person}{Chengming Li}, \bibinfo{person}{Xiaohai Xu}, {et~al\mbox{.}}} \bibinfo{year}{2021}\natexlab{b}.
\newblock \showarticletitle{Milvus: A purpose-built vector data management system}. In \bibinfo{booktitle}{\emph{SIGMOD}}. \bibinfo{pages}{2614--2627}.
\newblock


\bibitem[\protect\citeauthoryear{Wang, Lv, Xu, Wang, Yue, and Ni}{Wang et~al\mbox{.}}{2023}]%
        {wang2024efficient}
\bibfield{author}{\bibinfo{person}{Mengzhao Wang}, \bibinfo{person}{Lingwei Lv}, \bibinfo{person}{Xiaoliang Xu}, \bibinfo{person}{Yuxiang Wang}, \bibinfo{person}{Qiang Yue}, {and} \bibinfo{person}{Jiongkang Ni}.} \bibinfo{year}{2023}\natexlab{}.
\newblock \showarticletitle{An efficient and robust framework for approximate nearest neighbor search with attribute constraint}. In \bibinfo{booktitle}{\emph{NeurIPS}}, Vol.~\bibinfo{volume}{36}. \bibinfo{pages}{15738--15751}.
\newblock


\bibitem[\protect\citeauthoryear{Wang, Xu, Yue, and Wang}{Wang et~al\mbox{.}}{2021a}]%
        {wang2021comprehensive}
\bibfield{author}{\bibinfo{person}{Mengzhao Wang}, \bibinfo{person}{Xiaoliang Xu}, \bibinfo{person}{Qiang Yue}, {and} \bibinfo{person}{Yuxiang Wang}.} \bibinfo{year}{2021}\natexlab{a}.
\newblock \showarticletitle{A comprehensive survey and experimental comparison of graph-based approximate nearest neighbor search}. In \bibinfo{booktitle}{\emph{VLDB}}. \bibinfo{pages}{1964--1978}.
\newblock


\bibitem[\protect\citeauthoryear{Weaviate}{Weaviate}{2019}]%
        {weaviate}
\bibfield{author}{\bibinfo{person}{Weaviate}.} \bibinfo{year}{2019}\natexlab{}.
\newblock \bibinfo{booktitle}{\emph{Weaviate.io}}.
\newblock
\urldef\tempurl%
\url{https://weaviate.io/developers/weaviate/concepts/vector-index.}
\showURL{%
\tempurl}


\bibitem[\protect\citeauthoryear{Wei, Wu, Wang, Lou, Zhan, Li, and Cai}{Wei et~al\mbox{.}}{2020}]%
        {wei2020analyticdb}
\bibfield{author}{\bibinfo{person}{Chuangxian Wei}, \bibinfo{person}{Bin Wu}, \bibinfo{person}{Sheng Wang}, \bibinfo{person}{Renjie Lou}, \bibinfo{person}{Chaoqun Zhan}, \bibinfo{person}{Feifei Li}, {and} \bibinfo{person}{Yuanzhe Cai}.} \bibinfo{year}{2020}\natexlab{}.
\newblock \showarticletitle{AnalyticDB-V: a hybrid analytical engine towards query fusion for structured and unstructured data}.
\newblock \bibinfo{journal}{\emph{Proceedings of the VLDB Endowment}} \bibinfo{volume}{13}, \bibinfo{number}{12} (\bibinfo{year}{2020}), \bibinfo{pages}{3152--3165}.
\newblock


\bibitem[\protect\citeauthoryear{Wu, He, Qiao, Fu, Liu, and Yu}{Wu et~al\mbox{.}}{2022}]%
        {wu2022hqann}
\bibfield{author}{\bibinfo{person}{Wei Wu}, \bibinfo{person}{Junlin He}, \bibinfo{person}{Yu Qiao}, \bibinfo{person}{Guoheng Fu}, \bibinfo{person}{Li Liu}, {and} \bibinfo{person}{Jin Yu}.} \bibinfo{year}{2022}\natexlab{}.
\newblock \showarticletitle{HQANN: Efficient and robust similarity search for hybrid queries with structured and unstructured constraints}. In \bibinfo{booktitle}{\emph{CIKM}}. \bibinfo{pages}{4580--4584}.
\newblock


\bibitem[\protect\citeauthoryear{Xu, Li, Wang, and Xia}{Xu et~al\mbox{.}}{2020}]%
        {xu2020multiattribute}
\bibfield{author}{\bibinfo{person}{Xiaoliang Xu}, \bibinfo{person}{Chang Li}, \bibinfo{person}{Yuxiang Wang}, {and} \bibinfo{person}{Yixing Xia}.} \bibinfo{year}{2020}\natexlab{}.
\newblock \showarticletitle{Multiattribute approximate nearest neighbor search based on navigable small world graph}.
\newblock \bibinfo{journal}{\emph{Concurrency and Computation: Practice and Experience}} \bibinfo{volume}{32}, \bibinfo{number}{24} (\bibinfo{year}{2020}), \bibinfo{pages}{e5970}.
\newblock


\bibitem[\protect\citeauthoryear{Yahoo}{Yahoo}{2016}]%
        {NGTonline}
\bibfield{author}{\bibinfo{person}{Yahoo}.} \bibinfo{year}{2016}\natexlab{}.
\newblock \bibinfo{booktitle}{\emph{Nearest neighbor search with neighborhood graph and tree for high-dimensional data}}.
\newblock
\urldef\tempurl%
\url{https://github.com/yahoojapan/NGT.}
\showURL{%
\tempurl}


\bibitem[\protect\citeauthoryear{Yang, Li, Fang, and Wei}{Yang et~al\mbox{.}}{2020}]%
        {yang2020pase}
\bibfield{author}{\bibinfo{person}{Wen Yang}, \bibinfo{person}{Tao Li}, \bibinfo{person}{Gai Fang}, {and} \bibinfo{person}{Hong Wei}.} \bibinfo{year}{2020}\natexlab{}.
\newblock \showarticletitle{Pase: Postgresql ultra-high-dimensional approximate nearest neighbor search extension}. In \bibinfo{booktitle}{\emph{SIGMOD}}. \bibinfo{pages}{2241--2253}.
\newblock


\bibitem[\protect\citeauthoryear{Yianilos}{Yianilos}{1993}]%
        {yianilos1993data}
\bibfield{author}{\bibinfo{person}{Peter~N Yianilos}.} \bibinfo{year}{1993}\natexlab{}.
\newblock \showarticletitle{Data structures and algorithms for nearest neighbor search in general metric spaces}. In \bibinfo{booktitle}{\emph{SODA}}. \bibinfo{pages}{311--321}.
\newblock


\bibitem[\protect\citeauthoryear{Zhang, Yao, Gao, Wu, He, Li, Lu, Zhan, and Tang}{Zhang et~al\mbox{.}}{2023}]%
        {zhang2023learning}
\bibfield{author}{\bibinfo{person}{Pengcheng Zhang}, \bibinfo{person}{Bin Yao}, \bibinfo{person}{Chao Gao}, \bibinfo{person}{Bin Wu}, \bibinfo{person}{Xiao He}, \bibinfo{person}{Feifei Li}, \bibinfo{person}{Yuanfei Lu}, \bibinfo{person}{Chaoqun Zhan}, {and} \bibinfo{person}{Feilong Tang}.} \bibinfo{year}{2023}\natexlab{}.
\newblock \showarticletitle{Learning-based query optimization for multi-probe approximate nearest neighbor search}.
\newblock \bibinfo{journal}{\emph{The VLDB Journal}} \bibinfo{volume}{32}, \bibinfo{number}{3} (\bibinfo{year}{2023}), \bibinfo{pages}{623--645}.
\newblock


\bibitem[\protect\citeauthoryear{Zheng, Xi, Weng, Hung, Liu, and Jensen}{Zheng et~al\mbox{.}}{2020}]%
        {zheng2020pm}
\bibfield{author}{\bibinfo{person}{Bolong Zheng}, \bibinfo{person}{Zhao Xi}, \bibinfo{person}{Lianggui Weng}, \bibinfo{person}{Nguyen Quoc~Viet Hung}, \bibinfo{person}{Hang Liu}, {and} \bibinfo{person}{Christian~S Jensen}.} \bibinfo{year}{2020}\natexlab{}.
\newblock \showarticletitle{PM-LSH: A fast and accurate LSH framework for high-dimensional approximate NN search}. In \bibinfo{booktitle}{\emph{VLDB}}. \bibinfo{pages}{643--655}.
\newblock


\bibitem[\protect\citeauthoryear{Zuo and Deng}{Zuo and Deng}{2023}]%
        {zuo2023arkgraph}
\bibfield{author}{\bibinfo{person}{Chaoji Zuo} {and} \bibinfo{person}{Dong Deng}.} \bibinfo{year}{2023}\natexlab{}.
\newblock \showarticletitle{ARKGraph: All-Range Approximate K-Nearest-Neighbor Graph}. In \bibinfo{booktitle}{\emph{VLDB}}. \bibinfo{pages}{2645--2658}.
\newblock


\bibitem[\protect\citeauthoryear{Zuo, Qiao, Zhou, Li, and Deng}{Zuo et~al\mbox{.}}{2024}]%
        {zuo2024serf}
\bibfield{author}{\bibinfo{person}{Chaoji Zuo}, \bibinfo{person}{Miao Qiao}, \bibinfo{person}{Wenchao Zhou}, \bibinfo{person}{Feifei Li}, {and} \bibinfo{person}{Dong Deng}.} \bibinfo{year}{2024}\natexlab{}.
\newblock \showarticletitle{SeRF: Segment Graph for Range-Filtering Approximate Nearest Neighbor Search}. In \bibinfo{booktitle}{\emph{SIGMOD}}. \bibinfo{pages}{1--26}.
\newblock


\end{thebibliography}
\end{document}